\documentclass[twocolumn,showpacs,preprintnumbers,amsmath,amssymb]{revtex4}
\usepackage{graphicx,psfrag}% Include figure files
\usepackage{amssymb}
\usepackage{amsmath}
\usepackage[squaren,Gray,textstyle]{SIunits}
\usepackage{dcolumn}% Align table columns on decimal point
\usepackage{bm}% bold math

\def\nc{\newcommand}

\def\lsim{\mathrel{\raise.3ex\hbox{$<$\kern-.75em\lower1ex\hbox{$\sim$}}}}
\def\gsim{\mathrel{\raise.3ex\hbox{$>$\kern-.75em\lower1ex\hbox{$\sim$}}}}
\nc{\half}{\frac{1}{2}}
\nc{\shalf}{\ensuremath{\textstyle \frac{1}{2}}}

\nc{\deldag}{\mathbin{\partial\mkern-10.5mu\big/}}
\nc{\kdag}{\mathbin{k\mkern-10mu\big/}}
\nc{\Pdag}{\mathbin{P\mkern-10mu\big/}}

\nc{\beq} {\begin{equation}}
\nc{\eeq} {\end{equation}}
\nc{\beqa}{\begin{eqnarray}}
\nc{\eeqa}{\end{eqnarray}}

\begin{document}

\preprint{APS/123-QED}

\title{Dark Energy, Scalar-Tensor Gravity and Large Extra Dimensions}

\author{Kimmo Kainulainen}
   \email{Kimmo.Kainulainen@phys.jyu.fi}
\author{Daniel Sunhede}
   \email{Daniel.Sunhede@phys.jyu.fi}
\affiliation{
    Dept.~of Physics, University of Jyv\"askyl\"a, \\
    P.O.Box 35 (YFL),  FIN-40014 University of Jyv\"askyl\"a
}

\date{\today}% It is always \today, today.

\begin{abstract}
We explore in detail a dilatonic scalar-tensor theory of gravity inspired by large extra dimensions, where a radion field from compact extra dimensions gives rise to quintessence in our 4-dimensional world. We show that the model can give rise to other types of cosmologies as well, some more akin to $k$-essence and possibly variants of phantom dark energy. In our model the field (or radius) stabilization arises from quantum corrections to the effective 4D Ricci scalar. We then show that various constraints nearly determine the model parameters, and give an example of a quintessence-type cosmology consistent with observations. We show that the upcoming SNAP-experiment would easily distinguish the present model from a constant $\Lambda$ model with an \emph{equal} amount of dark energy, but that the SNAP-data alone will not be able distinguish it from a $\Lambda$ model with about 5\% less dark energy.
\end{abstract}

% Cosmology, 98.80.-k
% Particle-theory and field-theory models of the early Universe, 98.80.Cq
% Dark matter (stellar, interstellar, galactic, and cosmological), 95.35.+d
% Gravity in more than four dimensions, KaluzaÐKlein theory, unified field
%    theories; alternative theories of gravity, 04.50.+h 

\maketitle

%%%%%%%%%%%%%%%%%%%%%%%%%%%%%%%%%%%%%%%%%%%%%%%%%%%%%%%%%%%%%%%%%%%%%%%%%%
%
%  --------------------------- Section 1 -------------------------------
%
%%%%%%%%%%%%%%%%%%%%%%%%%%%%%%%%%%%%%%%%%%%%%%%%%%%%%%%%%%%%%%%%%%%%%%%%%%

\section{Introduction}
\label{sect:Intro}

The discovery that the expansion of the universe is accelerating implies the existence of an extremely small, but non-zero vacuum energy density. In summary, the observations have led to a new cosmological concordance model, according to which~\cite{wmap} $\Omega_{\rm {tot}} = 1.02 \pm 0.02 $, $\Omega_{\Lambda} \approx 0.73$, and $\Omega_{\rm {m}} \approx 0.27$. That is, the universe is flat and dominated by a ``dark energy'' component $\Omega_{\Lambda}$ which has an equation of state close to that of a bare cosmological constant today. The physical nature of the dark energy component is not known, but according to the popular \emph{quintessence} scenario~\cite{peebles88, ratra88, wetterich88}, it corresponds to energy stored into a classical background scalar field with some very particular properties. An interesting class of quintessence models can be characterized by a Lagrangian~\cite{ratra88, wetterich88, exp_1-7, albrecht00} %
\begin{equation} \label{eq:LNquint}
    \mathcal{L} = \frac{1}{2} (\partial \chi)^2
    - V_{\rm p}(\chi) e^{-\lambda \chi} \; ,
\end{equation}
where $V_{\rm p}(\chi)$ is some power-law modification to the dominant exponential factor. A desirable feature of such exponential potentials is that a small change in the field value $\chi$ can cause a large change in the energy density $\rho_{\chi}$. Moreover, they often allow scaling solutions, where the kinetic and potential energies of $\chi$ maintain a fixed ratio, leading to a constant equation-of-state parameter $w_{\chi}$ and to $\rho_{\chi}$ scaling exactly as a power of the scale factor. This in particular makes it possible for the quintessence field to mimic a cosmological constant.

Generic quintessence scenarios often rely on an \emph{ad.~hoc.} quintessence field with some suitably adjusted potential. A more complete physical theory should explain the origin of the --- possibly effective --- quintessence field and naturally predict the exponential form of its potential. Remarkably, these requirements can be met in a model with large extra dimensions, studied by Albrecht, Burgess, Ravndall, and Skordis (ABRS)~\cite{albrecht02_1, albrecht02_2, burgess02}. In the ABRS scenario the quintessence field $\chi$ corresponds to the radion mode, which after a compactification of the originally 6-dimensional space-time manifold shows up as an effective Jordan-Brans-Dicke field in the residual 4-dimensional action. A crucial ingredient of the ABRS scenario is its introduction of a scalar field $\phi$ living in the 6D bulk space, causing two key features that make the model special: First, the zero-point fluctuation of the bulk scalar field $\phi$ induce a Casimir potential $U_{\rm C} \sim U_0/r^4$ for the radion, which eventually acts as an effective cosmological constant. Second, the bulk scalar field $\phi$ also induce logarithmic corrections to both kinetic terms and the radion potential~\cite{kantowski87, birmingham88}, which are crucial for the radius stabilization mechanism. The proper amount of dark energy can arise in a parametrically natural way from the Casimir potential $U_{\rm C}$. In a model with $d$ equally large extra dimensions, the effective strength of gravity is $M_{\rm b}^{2+d}r^d$. Hence, since the naturalness condition $U_0 \sim 1$ requires a radius $r\sim (10^{-3} {\rm eV})^{-1}$ to yield the desired amount of dark energy, a 6D model corresponds to electroweak scale $M_{\rm b} \sim 1$ TeV, whereas for $d=3$ one would need a lower scale $M_{\rm b} \sim 1$ GeV. Seen the other way around, setting the fundamental scale to TeV in a model with $d>2$, would require fine-tuning $U_0$ to extremely small values. In this sense the 6D case is indeed special, since it naturally yields both the electroweak scale and the desired amount of dark energy. After obtaining the general form of the action in section \ref{sect:model}, we will set $d = 2$ for the remainder of the paper.

We will in this paper explore a class of dilatonic scalar-tensor theories of gravity (STG models for short) inspired by the ABRS scenario. The above extra-dimensional models can indeed be recast as a dilatonic STG model by simply defining the field $\varphi \equiv \log{M_{\rm b} r}$.  Similar to theories with large extra dimensions, a dilatonic factor $e^{\lambda \varphi}$ allows the fundamental scale of gravity $M_{\rm b}$ to be much lower than $M_{\rm Pl}$. Moreover, when written in the Einstein frame, any dilatonic STG model will introduce an exponential behaviour to the potential of the theory. In a general scalar-tensor theory, not only the scalar field potential can cause an observed acceleration of the universe, but also the functions associated with the kinetic terms of the theory. This paper will however focus on the possibility of an accelerated expansion originating in the potential term. Although the connection to the ABRS scenario provides additional motivation, it should be stressed that our results are independent of origin and the presented scalar-tensor theory can indeed be viewed as the starting point.

The present model is strongly confronted by the existing astrophysical constraints, and most notably by precision tests of General Relativity in the solar system~\cite{will98}. GR-constraints were already considered by the authors of ref.~\cite{albrecht02_2}, but our results differ from theirs; in fact we found that they rule out the original scenario when correctly accounted for. The problem arises from the fact that the perturbative loop corrections coming from the bulk scalar field $\phi$, which were treated as small quantities in ref.~\cite{albrecht02_2}, actually must be of order unity to fulfil the observational constraints. This is the case in particular for the scaling function $A(\varphi)$ for the effective 4D Ricci scalar $R_{\hat g}$ in the Jordan frame: At tree level $A(\varphi) = 1$ whereas the GR-constraint require that $A(\varphi) \approx 0$ today. Although this poses no problem when one takes the STG model of Eqn.~(\ref{eq:S}) as a pure phenomenological starting point, we point out that it is indeed also possible to fulfil this requirement in the context of a converging perturbation theory within the extra dimensional motivation. This is technically possible if the loop corrections come from \emph{many} independent very weakly coupled bulk fields $\phi_i$. The sacrifice in this scheme is that the original radius stabilization mechanism based on the Casimir potential does not work.

However, we have found that our model allows a new stabilization mechanism, based on the $\varphi$-scaling of the effective 4D Ricci scalar. Indeed, 
suppose that the function $A(\varphi)$ has a zero or becomes very small for some $\varphi = \varphi_*$. It turns out that a root $A(\varphi_*) = 0$ shows up as a singularity in the effective potential in the Einstein frame, that confines the field (and hence the radius in the extra-dimensional context) to the range $\varphi < \varphi_*$. We shall see that this mechanism works in the simplest perturbative parametrization of the loop-corrected action. However, the mechansim is much more generic; we can state that \emph{in any model where $A(\varphi)$ has a zero, for some large value $\varphi=\varphi_*$, the effective potential becomes confining and develops a minimum close to $\varphi_*$, where $A(\varphi_{\rm min}) \approx 0$.}  If strange at first sight, it should be 
noted that this scaling arises naturally in the present model, and it is in fact all but a {\it necessary} condition for satisfying the GR-constraint at present. Conversely, in the context of this stabilation scheme the GR-constraint becomes almost automatically satisfied.

Astrophysical and theoretical constraints nearly fix the relevant model parameters. Moreover, for a given set of parameters, there is practically a unique set of initial conditions that can give rise to correct values for the cosmological parameters today. We present a particular quintessence-like solution which obeys all constraints, starting from time before nucleosynthesis until today. We then argue that our model makes a generic prediction for the evolution of the equation-of-state parameter $w_{\rm tot}(z)$, which in principle makes it distinguishable from a flat, constant $\Lambda$ cosmology. We then show that the sensitivity of the forthcoming SNAP-experiment~\cite{SNAP} should be good enough to separate our model from a constant $\Lambda$ model with \emph{equal} amount of dark energy today. However, the SNAP-data~\cite{SNAPdata} alone will not be able to break a degeneracy between our model and a $\Lambda$ model with slightly different amount of dark energy (about 5\% less).

While our scalar-tensor action is very simple in the Jordan frame, it introduces a very complicated structure for the kinetic term of the scalar field $\varphi$ when written in the Einstein frame. Even with the little room one has to play after applying the constraints, it is in fact possible to arrange model parameters such that this term becomes very large and possibly singular, or even negative. In either of the former cases the model classifies rather as a $k$-essence than a quintessence scenario, and in the latter it gives rise to variants of recently introduced phantom cosmologies~\cite{phantom}. In this paper we mostly concentrate on non-phantom models. While we do give some examples of parametrizations leading to phantom cosmologies, it is not yet clear to us whether it is possible to make a transition into such a phase and make the scenario compatible with observations.

The paper is organized as follows. We begin in section \ref{sect:model} by introducing the scalar-tensor theory, and show its motivation within extra-dimensional theories, by reviewing the basic ABRS scenario and its reduction to an effective 4D gravity theory. We continue in section \ref{sect:k-essence} by discussing how the effective 4D theory relates to other dark energy models, including $k$-essence, quintessence, and phantom cosmology. We also derive an effective action for the model in terms of the field $\chi$ in the Einstein frame, and from it the coupled cosmological evolution equations for $\chi$ and various density components. Section \ref{sec:newstab} begins with a discussion of a problem with the perturbative expansion employed by ABRS and then outlines a perturbatively reliable scheme with a new mechanism for the radius stabilization. Having obtained a consistent scenario, we continue in section \ref{sect:constraints} by deriving constraints on the model parameters, arising from various theoretical and astrophysical considerations. In section \ref{sect:realistCosm} we then present a realistic cosmological evolution from times before nucleosynthesis until present that satisfies all observational constraints. We also compute the prediction of the model for the supernovae magnitude-redshift relation and compare it with cosmological constant models. Finally, section \ref{sect:summary} contains our summary and the conclusions.

% New Section
%%%%%%%%%%%%%%%%%%%%%%%%%%%%%%%%%%%%%%%%%%%%%%%%%%%%%%%%%%%%%%%%%%%%%%%%%%
%
%  --------------------------- Section 2 ------------------------------
%
%%%%%%%%%%%%%%%%%%%%%%%%%%%%%%%%%%%%%%%%%%%%%%%%%%%%%%%%%%%%%%%%%%%%%%%%%%

\section{The Model}
\label{sect:model}

The aim of this paper is to explore the alternative cosmologies of a class of scalar-tensor theories of gravity described by an effective action
\begin{widetext}
\begin{equation}
   S  =  \int {\rm d}^4 x \sqrt{-\hat{g}}
      \left[
         \frac{M_{\rm b}^2 e^{\lambda\varphi}}{2}
            \Big( A(\varphi) R_{\hat{g}} -
               2 B(\varphi) (\partial \varphi)^2
            \Big)
         - U(\varphi) + \hat{\mathcal{L}}_{\rm m}
      \right] \: .
\label{eq:S}
\end{equation}
\end{widetext}
%
% \begin{eqnarray}
%    S  &=&  \int {\rm d}^4 x \sqrt{-\hat{g}} \Bigg[
%          \frac{M_{\rm b}^2 e^{\lambda\varphi}}{2}
%             \Big( A(\varphi) R_{\hat{g}} \\Ê\nonumber            
%       & & \phantom{\int {\rm d}^4 x \sqrt{-\hat{g}} \Bigg[}
%          {}- 2B(\varphi) (\partial \varphi)^2 \Big)
%          - U(\varphi) + \hat{\mathcal{L}}_{\rm m}
%       \Bigg] \: .
% \label{eq:S}
% \end{eqnarray}
%
Here $M_{\rm b}$ is the fundamental scale of gravity, $R_{\hat{g}}$ is the Ricci scalar of the metric $\hat{g}_{\mu\nu}$, and $\hat{\mathcal{L}}_{\rm m}$ is the classsical Lagrangian of all matter contained within the universe. The dilatonic factor $e^{\lambda \varphi}$ allows the scale $M_{\rm b}$ to be much lower than $M_{\rm Pl}$, but apart from this behaviour, $A(\varphi)$ and $B(\varphi)$ can be arbitrary functions. We will however see that the theory gives rise to interesting cosmologies already for the simple case when $A(\varphi)$ and $B(\varphi)$ are linear in $\varphi$. One of the two functions $A(\varphi)$ and $B(\varphi)$ can of course be set to 1 by a redefinition of $\varphi$, but we have kept the above form since it will make the connection to extra-dimensional scenarios more explicit. While the detailed form of $U(\varphi)$ is not constrained in a generic STG model, we will set $U(\varphi) \propto e^{-4\varphi}$, since this will connect our class of scalar-tensor theories to the brane-world scenario to be discussed below. Note that for $A(\varphi) = B(\varphi) = 1$ and $U(\varphi) = 0$ the action (\ref{eq:S}) corresponds to a basic Jordan-Brans-Dicke theory with coupling parameter $\omega = 2/\lambda^2$.

The above scalar-tensor theory, Eqn.~(\ref{eq:S}), can be viewed as a starting point of this paper, but is should be stressed that the form of the action is motivated by extra-dimensional scenarios~\cite{albrecht02_1,albrecht02_2}. Because this will also lead to a more specific prediction of the form of the functions $A$, $B$ and $U$, the present model is potentially more predictive than previous studies of scalar-tensor gravity and quintessence, where the scalar field corresponded to a dilaton from string theory~\cite{dilatonquint, DPV}, or the origin of the field was not considered in depth~\cite{STGquint, extended}. Also, in the present model, all mass scales and parameters have perfectly natural values and no fine-tuning is needed for ending up at the desired amount of dark energy at present.

\subsection{The Brane World}

Let us consider a brane world scenario in which our $(3+1)$-dimensional space is a brane embedded in a $(4+d)$-dimensional bulk space. All standard model fields are confined to the four-dimensional brane while gravity propagates in all dimensions. It is well known that if the size of the extra dimensions $r$ is large, the scale of physics on the brane $M_{\rm b}$ can be of electroweak scale, hence providing a compelling solution to the hierarchy problem~\cite{Hamedetal, Antoniadetal}. The effective 4D gravitational coupling in this scenario is $(8 \pi G_{\rm eff})^{-1/2} \equiv M_{\rm b} (M_{\rm b}r)^{d/2} \sim M_{\rm Pl} \sim 10^{18}$ GeV, corresponding to a very large dimensionless parameter $M_{\rm b}r \sim 10^{30/d}$ TeV$/M_{\rm b}$.

The full action of the theory can be divided into two terms, $S = S_{\rm B} + S_{\rm b}$, corresponding to the bulk and the brane contributions respectively. The leading part of the bulk action is
\begin{equation}
   S_{\rm B} =
      \frac{M_{\rm b}^{2+d}}{2} \int {\rm d}^{4}x {\rm d}^{d}y
         \sqrt{\textrm{\footnotesize{(}$-$\footnotesize{)}}^{d+1}
            \mathcal{G} \phantom{.}}
         \mathcal{R} + \ldots \: ,
\label{eq:SBulk}
\end{equation}
where $\mathcal{R}$ is the Ricci scalar of the $(4+d)$-dimensional metric $\mathcal{G}_{MN}$ and the ellipses represent possible bulk fields. The detailed form of the brane action is not important; the crucial part is that it only depends on the \emph{four}-dimensional metric $\hat{g}_{\mu \nu}$:
\begin{equation}
    S_{\rm b} = \int {\rm d}^{4}x \sqrt{-\hat{g}}
        \hat{\mathcal{L}}_{\rm m} (\hat{g}_{\mu \nu}, \ldots) \: .
\label{eq:Sbrane}
\end{equation}
All matter fields, denoted by ellipses in $\hat{\mathcal{L}}_{\rm m}$, are trapped on the brane and do not carry stress-energy into the extra dimensions. The bare cosmological constant $\Lambda$ is here assumed to vanish; at present our model does not offer any insight as to how this is accomplished. Further, we assume that the $(4+d)$-dimensional metric has a block-diagonal structure:
\begin{equation}
  \mathcal{G}_{MN} =
    \left[
       \begin{array}{cc}
         \hat{g}_{\mu \nu}(x) & 0 \\
           0  & \varrho^{2}(x) h_{mn}(y)
       \end{array}
    \right] \: .
\label{eq:GMN}
\end{equation}
That is, the $(4+d)$-dimensional space-time manifold consists of a four-dimensional brane with metric $\hat{g}_{\mu\nu}$, connected to a $d$-dimensional surface with metric $h_{mn}$, such that the size of the extra dimensions, $\varrho (x) = M_{\rm b}r(x)$, is allowed to vary as a function of the position on the brane.

\subsection{Effective 4D Theory}    
\label{sect:4DEffect}

We are obviously interested in the effective four dimensional theory on the brane at large scales. It can be obtained from Eqns.~(\ref{eq:SBulk}--\ref{eq:GMN}) via standard dimensional reduction. Indeed, integrating out the extra spatial dimensions at the tree-level one finds the effective action:
\begin{widetext}
\begin{equation}
   S_{\rm eff}  =  \int {\rm d}^4 x \sqrt{-\hat{g}}
      \left[
         \frac{M_{\rm b}^2 (M_{\rm b}r)^d}{2}
            \left(  R_{\hat{g}}
            - d(d-1) \left( \frac{\partial r}{r} \right)^2
            + \frac{M_{\rm b}^d}{(M_{\rm b}r)^2} \int {\rm d}^{d}y
               \sqrt{\textrm{\footnotesize{(}$-$\footnotesize{)}}^{d}h} R_h
            \right)
         + \hat{\mathcal{L}}_{\rm m}(\hat{g})
      \right] \: ,
\label{eq:Seff}
\end{equation}
\end{widetext}
where $R_{\hat{g}}$ and $R_{h}$ represent the Ricci scalars of $\hat{g}_{\mu \nu}$ and $h_{mn}$, respectively, and the normalization of $h_{mn}$ is chosen such that $M_{\rm b}^d \int {\rm d}^{d}y \sqrt{\textrm{\footnotesize{(}$-$\footnotesize{)}}^{d}h} \equiv 1$.

The action (\ref{eq:Seff}) is a modification of the regular Einstein Hilbert theory of gravity, and it can be recast into a dilatonic scalar-tensor model by a change of variables $\varphi \equiv \log{M_{\rm b} r}$. It is well known that the presence of such a scalar field will yield long range forces in addition to the regular gravitational force originating in $\hat{g}_{\mu \nu}$. As we will see in section \ref{sect:constraints}, precision measurements of general relativity on the solar system scale impose a strong constraint $\alpha \approx 0$, where $\alpha$ is the coupling strength of the scalar field to matter (see Eqn.~(\ref{eq:alpha}) below). In the tree level action (\ref{eq:Seff}) we would obtain $\alpha^2 = d/(d+2)$, so that  $\alpha \sim 1$ at tree level in a generic model with d compactified large extra dimensions.

It is therefore essential that the action can be modified from its tree level form. If for example $(M_{\rm b}r)^d \to (M_{\rm b}r)^d A$, then $\alpha$ can be small if ${\rm d}A/{\rm d}\varphi + dA \approx 0$ for large $M_{\rm b}r$, where $\varphi \equiv \log{M_{\rm b} r}$. A simple example of this is a case when $A$ is a linear function with a zero in $\log{M_{\rm b} r}$. Remarkably, such logarithms are generically induced by quantum corrections from bulk fields $\phi_i$ living in large extra dimensions~\cite{kantowski87, birmingham88,albrecht02_1}. These include corrections to the kinetic terms, in particular an $A$-term as requested, but also a Casimir potential $U_{\rm C}$, which for toroidal compactification has the leading term~\cite{albrecht02_1}
\begin{equation} \label{eq:Ucasimir}
    U_{\rm C} = \frac{U_0}{r^{4}} + \ldots \: .
\end{equation}
The Casimir potential also receives logaritmic quantum corrections.

The curvature term proportional to $R_h$ in Eqn.~(\ref{eq:Seff}) is non-vanishing except for Ricci flat compactifications. Indeed, if $R_h \sim 1/M_{\rm b}^2$, this term would be too large to be consistent with the observed amount of the dark energy and the value of Newtons constant for any number of extra dimensions. We are therefore led to consider only topologies where the contribution from $R_h$ vanish, such as the trivial case when the extra dimensions are Ricci flat ($R_h = 0$)~\footnote{It is interesting to note that for brane worlds in 6D supergravity, the bulk curvature is cancelled by the brane tensions~\cite{SLEDtension}. For further connection of the present model to 6D supersymmetric large extra dimensions see also ref.~\cite{SLED}.}. These considerations lead us to parametrize the action of the effective 4D theory as follows:
\begin{widetext}
\begin{equation}
   S_{\rm eff}  =  \int {\rm d}^4 x \sqrt{-\hat{g}}
      \left[
         \frac{M_{\rm b}^2 (M_{\rm b}r)^d}{2}
            \left( A(r) R_{\hat{g}} -
               B(r)d(d-1) \left( \frac{\partial r}{r} \right)^2
            \right)
         - C(r)\frac{U_0}{r^4} + \hat{\mathcal{L}}_{\rm m}(\hat{g})
      \right] \: ,
\label{eq:SJordan}
\end{equation}
\end{widetext}
where
\begin{eqnarray}
  A(r) &\approx& 1 + a \log{M_{\rm b} r} \: ,
  \label{eq:A} \\
  B(r) &\approx& 1 + b \log{M_{\rm b} r} \: ,
\label{eq:B} \\
  C(r) &\approx& 1 + c \log{M_{\rm b} r} \: ,
\label{eq:C}
\end{eqnarray}
and $a$, $b$, and $c$ are unknown but small parameters proportional to some dimensionless coupling constants of the theory.

Let us now return to the original scalar-tensor theory. Indeed, comparing the action (\ref{eq:S}) with the effective 4D action of the brane-world scenario, Eqn.~(\ref{eq:SJordan}), we see that they are completely equivalent, and can be put into identical form via the definitions
\begin{equation}
    \varphi \equiv \log{M_{\rm b} r} \: ,
\label{eq:phimbr}
\end{equation}
\begin{equation}
    U(\varphi) \equiv C(\varphi) U_0 e^{-4\varphi} \: ,
\label{eq:U}
\end{equation}
and $A(\varphi) = 1 + a\varphi$, $B(\varphi) =1 + b\varphi$, $C(\varphi) = 1 +c\varphi$. Here, $\lambda = d$ and we now fix the number of extra dimensions to $d = 2$. For this particular choice of $d$, the naturalness condition $U_0 \sim 1$ and the desire to obtain the observed amount of dark energy corresponds to the electroweak scale $M_b \sim 1$ TeV. Indeed, this explicit form of the action will be the focus for the remainder of the paper. Note that we treat all fields in the action (\ref{eq:SJordan}), and hence $\varphi$, as purely classical fields. The corrections in the form of a Casimir potential and the coefficients $A$, $B$, and $C$ arise from the presence of quantized bulk fields $\phi_i$.

The parametrization of the model (\ref{eq:SJordan}) was first derived by Albrecht, Burgess, Ravndal and Skordis, in the context of a six-dimensional model with a bulk scalar field $\phi$ with renormizable cubic self-interaction and a toroidal compactification~\cite{albrecht02_1, albrecht02_2}. It has later been realized~\cite{SLEDcorr} that logarithmic corrections actually \emph{do not} arise for this simple toroidal compactification and boundary 
conditions. However, logarithmic corrections from bulk fields $\phi_i$ have been found to be a generic outcome for a variety of compactification schemes~\cite{kantowski87, birmingham88} and moreover, it is likely that they are generated also for torii given twisted boundary conditions and/or orbifold symmetries~\cite{SLEDcorr}. We thus believe that the effective actions (\ref{eq:S}) and (\ref{eq:SJordan}) have strong motivation based on large extra dimensional theories. Moreover, note that we basically derived the form (\ref{eq:SJordan}) out of necessity: In a generic model with large extra dimensions a large curvature term $\propto R_h$ must be absent and the constraints from general relativity force us to introduce a scaling function $A$ such that $A \approx 0$ for large $M_{\rm b}r$. The action (\ref{eq:SJordan}) with the parametrizations (\ref{eq:A}--\ref{eq:C}) is perhaps the simplest one compatible with these requests, but as will be further discussed in section \ref{sect:k-essence}, the precise forms of the scaling functions are not important; it is enough if $A$ has a zero or just becomes very small for large $M_{\rm b}r$, i.e. the present value of $\varphi$.

\subsection{Casimir Radius Stabilization}
\label{sect:Casimirstab}

In the context of large extra dimensions (LED), the original ABRS argument for large radius stabilization of the compactified dimensions was based on the particular form of the quantum corrected Casimir potential~\cite{albrecht02_1, albrecht00, weller00} in Eqn.~(\ref{eq:SJordan}):
\begin{equation}
    U_{\rm C}(r) = U_0 M_{\rm b}^4 \left(\frac{1}{M_{\rm b} r}\right)^4
    \left[1 + c \log{M_{\rm b} r} \right] \: .
\label{eq:Udesired}
\end{equation}
$U_{\rm C}(r)$ always has a runaway minimum at $r \to \infty$. However, if $c$ is negative it develops another minimum at finite, but exponentially large  scale (in comparision with the microscopic scale $l = M_{\rm b}^{-1}$):
\begin{equation}
    M_{\rm b}r \approx \: \exp{\left(\frac{1}{4}
    + \frac{1}{|c|}\right)} \: .
\label{eq:largescale}
\end{equation}
Fixing $M_{\rm b}r$ to the desired value $\sim 10^{15}$ indeed sets the scale of radiative corrections to a small and perturbatively natural value $|c| \sim 1/50$. This simple argument neglects the coupling between the radius and the 4D metric in Eqn.~(\ref{eq:SJordan}) however. In reality one cannot make this assumption, and it will turn out that the other radiative corrections, and the $A$-term in particular, will play a crucial role for the stabilization in later developments.

Although the simplified line of reasoning above does not hold for the stabilization mechanism itself, it does set the right scale for the parameters $a$, $b$, and $c$ needed by a successfull cosmology. In the LED context the small scale of these parameters is natural since they arise as radiative corrections. One could imagine this to also be the case for the original scalar-tensor theory, Eqn.~(\ref{eq:S}), even if one chooses to neglect its connection to the brane-world scenario, and we will henceforth adopt $1/50$ as the natural scale for $a$, $b$, and $c$.

% New Section
%%%%%%%%%%%%%%%%%%%%%%%%%%%%%%%%%%%%%%%%%%%%%%%%%%%%%%%%%%%%%%%%%%%%%%%%%%
%
%  --------------------------- Section 3 ------------------------------
%
%%%%%%%%%%%%%%%%%%%%%%%%%%%%%%%%%%%%%%%%%%%%%%%%%%%%%%%%%%%%%%%%%%%%%%%%%%

\section{Quintessence, $k$-essence or Phantom Energy?}
\label{sect:k-essence}

We have so far written the model in the so called Jordan frame, where the metric $\hat{g}_{\mu \nu}$ is not canonically normalized. However, it is in the Jordan frame where all observations have their usual interpretation, since it is in this frame where the matter Lagrangian couples to the metric \emph{only} and hence making it possible to define standard clocks and rods that do not depend on the value of the scalar field. Nevertheless, the equations of motion are far more transparent in the Einstein frame (denoted below by a metric $g_{\mu \nu}$ without a hat), where the gravitational action has the more familiar form:
\begin{equation}
  \frac{M_{\rm b}^2 Ae^{\lambda \varphi}}{2} R_{\hat{g}}
    \equiv
    \frac{M_{\rm b}^2}{2} R_{g} + \textrm{additional terms} \: ,
\label{eq:Einstein-Jordan}
\end{equation}
where $\lambda = d$ and $R_g$ is the Ricci scalar of the Einstein metric $g_{\mu \nu}$. Let us define the commonly used variables
\begin{equation}
  F(\varphi) \equiv Ae^{d\varphi} \: , \qquad
  Z(\varphi) \equiv -d(d-1)Be^{d\varphi} \: ,
\label{eq:FZ}
\end{equation}
where, $F(\varphi)$ and $Z(\varphi)$ are the dimensionless functions multiplying the kinetic terms $R_{\hat{g}}$ and $(\partial \varphi)^2$, respectively, in the Jordan frame action (\ref{eq:SJordan}). Now, since schematically $R_g \propto g_{\mu \nu}^{-1}$, Eqn.~(\ref{eq:Einstein-Jordan}) implies the conformal transformation
\begin{equation}
  g_{\mu \nu} \equiv F(\varphi)\hat{g}_{\mu \nu} \: .
\label{eq:conformal}
\end{equation}
This recasts the action (\ref{eq:SJordan}) with our choice $d=2$, corresponding to the ABRS scenario, as
\begin{eqnarray}
   S_{\rm eff} &=& \int {\rm d}^4x {\sqrt{-g}}
      \Bigg[ \frac{M_{\rm b}^2}{2} R_g
         + \frac{1}{2} k^2(\varphi) (\partial \varphi)^2 \nonumber \\
      & & \phantom{\int {\rm d}^4x {\sqrt{-g}} \Bigg[}
         {}- V_{\rm E}(\varphi) + \mathcal{L}_{\rm m}(g, \varphi)
      \Bigg] \: ,
\label{eq:LEinsteinvarphi}
\end{eqnarray}
where
\begin{eqnarray}
  \frac{k^2(\varphi)}{M_{\rm b}^2} & \equiv & \frac{3}{2}
      \left(\frac{{\rm d} \log{F(\varphi)}}{{\rm d} \varphi}\right)^2
    + \frac{Z(\varphi)}{F(\varphi)} \nonumber \\
  & = & \frac{3}{2} \left(2 + \frac{a}{1+a\varphi}\right)^2
         - \frac{2(1+b\varphi)}{1+a\varphi} \: .
\label{eq:ksquared}
\end{eqnarray}
and
\begin{equation}
  V_{\rm E}(\varphi) \equiv \frac{U(\varphi)}{F^2(\varphi)}
     = M_{\rm b}^{4} U_0
        \frac{1+c\varphi}{(1+a\varphi)^2} e^{-8\varphi} \:
\label{eq:Vvarphi}
\end{equation}
As seen from Eqn.~(\ref{eq:Vvarphi}), any dilatonic theory with a non-zero $U(\varphi)$ will give rise to an exponential potential in the Einstein frame. Note that we have not expanded any of the above expressions in perturbative parameters; it will later become clear why this indeed is the sensible thing to do. Written in the form (\ref{eq:LEinsteinvarphi}) our effective theory has obvious resemblance with some popular toy models studied earlier in the literature. First, if the radiative corrections were neglected (or they are neglibly small so that effectively $a, b, c \simeq 0$) the potential reduces to a pure exponential and the kinetic energy becomes canonical apart from a constant scaling. In this limit our theory thus becomes very similar to the original \emph{quintessence} scenario~\cite{caldwell}. We shall see later that the model, despite t does lead to quintessence-like cosmological solutions.

In general, the function $k^2(\varphi)$ gives rise to a non-canonically normalized kinetic term for $\varphi$, which is the trademark of another class of dark energy models called \emph{$k$-essence}~\cite{HW2001,k-essence}. In particular our model bears similarity to mixed $k$-essence -- quintessence models discussed in ref.~\cite{HW2001}.  Observe in particular that in the Planck era, \emph{i.e.} when $\varphi \sim 0$ (corresponding to $r \sim M_{\rm b}^{-1}$ in the LED scenario) we have $k^2(\varphi) = 4M_{\rm b}^2$ and $V_{\rm E}(\varphi) = U_0 M_{\rm b}^4$. Thus the theory (\ref{eq:LEinsteinvarphi}) is natural (contains no arbitrarily small parameters) if only $U_0 \sim 1$. This is of course required independently by the naturalness of our underlying model from a field theory point of view, and we shall see later that it does give rise to acceptable cosmologies as well.

It is conceivable that for certain values of the parameters $a$ and $b$, the function $k^2(\varphi)$ becomes negative. In this case our model corresponds neither to $k$-essence nor to quintessence, but instead exhibits the features of \emph{phantom cosmologies}~\cite{phantom}, with an equation-of-state parameter $w_{\varphi} < -1$~\footnote{
   $w_{\varphi} \equiv p_{\varphi}/\rho_{\varphi} =
      \left(\frac{1}{2}k^2(\varphi)\dot{\varphi}^2 - V(\varphi)\right)
      / \left(\frac{1}{2}k^2(\varphi)\dot{\varphi}^2 + V(\varphi)\right)$.
   Strictly speaking, $\varphi$ is only a true phantom field (here defined as
   $w_{\varphi} < -1$) when
   $-V(\varphi) < \frac{1}{2}k^2(\varphi)\dot{\varphi}^2 < 0$.}.
(Of course, as was also pointed out in ref.~\cite{k-essence}, the phantom models could be viewed as forming a particular class of $k$-essence models.) For example, one could imagine that our model initially starts as quintessence or $k$-essence, and later in time changes its behaviour to a phantom cosmology, or vice versa. We shall show later such scenarios are indeed supported by natural values of the model parameters, but it is not yet clear to us whether they can be arranged dynamically in a model compatible with observations. For the major part of the paper we will however concentrate on the case where $k^2(\varphi) > 0$.

Let us finally note that the conformal transformation (\ref{eq:conformal}) between the Einstein and Jordan frame metrics is singular at $A(\varphi_*)=0$, and that changing the sign of $A$ will cause gravitons to carry negative energy in the Jordan frame and effectively reverse the signature of the Einstein metric. Perhaps fortunately this consideration appear mostly academic, as we shall see that it is practically impossible for any classical evolution starting from a $\varphi$ smaller that $\varphi_*$ to move past the singularity, hence keeping the field confined to values below $\varphi_*$ at all times.

\subsection{Non-Phantom Cosmologies}

When $k^2(\varphi) > 0$ we can make a further change of variables such that the kinetic term has the standard canonical form~\cite{HW2001}. In this case our theory belongs roughly to the class of $k$-essence and quintessence models. The difference between the latter two is that in $k$-essence the behaviour of the field is driven by its non-canonical kinetic term (in a pure $k$-essence the potential is actually vanishing~\cite{k-essence}), whereas in quintessence the desired behaviour follows from the special form of the potential. We will find that our model can at times behave as quintessence and at others as $k$-essence, as well as show some characteristics not typical to either of them. Indeed, introducing
\begin{equation}
\chi = K(\varphi) \qquad {\rm with} \qquad
k(\varphi) = \frac{{\rm d} K(\varphi)}{{\rm d} \varphi}
\label{eq:varphi-chi}
\end{equation}
we get the effective action
\begin{eqnarray}
   S_{\rm eff} &=& \int {\rm d}^4x {\sqrt{-g}}
      \Bigg[ \frac{M_{\rm b}^2}{2} R_g
         + \frac{1}{2} (\partial \chi)^2 \nonumber \\
      & & \phantom{\int {\rm d}^4x {\sqrt{-g}} \Bigg[}
         {}- V(\chi) + \mathcal{L}_{\rm m}(g, \chi)
      \Bigg] \: ,
\label{eq:LEinstein}
\end{eqnarray}
where $V(\chi) \equiv V_{\rm E}(\varphi) = V_{\rm E}(K^{-1}(\chi))$.

\begin{figure*}[!t] % X vs Mbr  and  V vs X figures
    \begin{center}
    \begin{psfrags}  % Begin psfrags sequence
    \psfragscanon%
    %
    % text strings Fig (a):
    \psfrag{f01}[t][t]{\setlength{\tabcolsep}{0pt}
      \begin{tabular}{c}Fig. (a)\end{tabular}}%
    \psfrag{s03}[b][b]{\setlength{\tabcolsep}{0pt}
        \begin{tabular}{c}$\chi$\end{tabular}}%
    \psfrag{s04}[t][t]{\setlength{\tabcolsep}{0pt}
        \begin{tabular}{c}$\varphi/\log{10}$\end{tabular}}%
    %
    % text strings Fig (b):
    \psfrag{f02}[t][t]{\setlength{\tabcolsep}{0pt}
      \begin{tabular}{c}Fig. (b)\end{tabular}}%
    \psfrag{s05}[t][t]{\setlength{\tabcolsep}{0pt}
        \begin{tabular}{c}$\chi$\end{tabular}}%
    \psfrag{s06}[b][b]{\setlength{\tabcolsep}{0pt}
        \begin{tabular}{c}$V(\chi)$\end{tabular}}%
    \psfrag{s07}[b][b]{\setlength{\tabcolsep}{0pt}
        \begin{tabular}{c}$\log{V(\chi)}$\end{tabular}}%
    %
    % X(Mbr) xticklabels:
    \psfrag{z01}[t][t]{$10$}%
    \psfrag{z02}[t][t]{$12$}%
    \psfrag{z03}[t][t]{$14$}%
    \psfrag{z04}[t][t]{$16$}%
    \psfrag{z05}[t][t]{$18$}%
    \psfrag{z06}[t][t]{$20$}%
    %
    % X(Mbr) yticklabels:
    \psfrag{y01}[r][r]{$-60$}%
    \psfrag{y02}[r][r]{$-40$}%
    \psfrag{y03}[r][r]{$-20$}%
    \psfrag{y04}[r][r]{$0$}%
    \psfrag{y05}[r][r]{$20$}%
    \psfrag{y06}[r][r]{$40$}%
    %
    % V(X) xticklabels:
    \psfrag{x01}[t][t]{$-30$}%
    \psfrag{x02}[t][t]{$-20$}%
    \psfrag{x03}[t][t]{$-10$}%
    \psfrag{x04}[t][t]{$0$}%
    \psfrag{x05}[t][t]{$10$}%
    \psfrag{x06}[t][t]{$20$}%
    %
    % V(X) yticklabels:
    \psfrag{v01}[r][r]{$0$}%
    \psfrag{v02}[r][r]{$0.2$}%
    \psfrag{v03}[r][r]{$0.4$}%
    \psfrag{v04}[r][r]{$0.6$}%
    \psfrag{v05}[r][r]{$0.8$}%
    \psfrag{v06}[r][r]{$1$}%
    \psfrag{v07}[r][r]{$1.2$}%
    \psfrag{ypower1}[Bl][Bl]{$\times 10^{-102}$}%
    %
    % log V(X) yticklabels:
	\psfrag{v11}[l][l]{$-300$}%
	\psfrag{v12}[l][l]{$-290$}%
	\psfrag{v13}[l][l]{$-280$}%
	\psfrag{v14}[l][l]{$-270$}%
	\psfrag{v15}[l][l]{$-260$}%
	\psfrag{v16}[l][l]{$-250$}%
	\psfrag{v17}[l][l]{$-240$}%
	\psfrag{v18}[l][l]{$-230$}%
    %
    % Figure:
    \includegraphics[width=15.4cm]{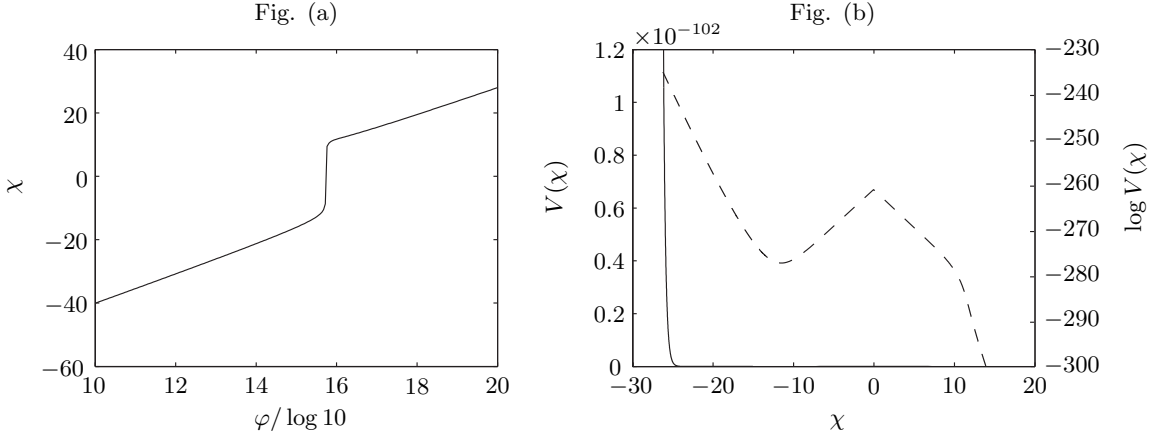}%
    \end{psfrags}%
    \end{center}
    \caption{\small{(a) $\chi$ in terms of $\varphi/\log{10}$
        ($= \log_{10}{M_{\rm b} r}$) for the representative case
        $a = -0.0276$ and $b = -0.031$. (b) The potential $V(\chi)$
        for the same $a$ and $b$ with $U_0=2.60$ and $c = 0$.
        The dashed line represents $\log{V(\chi)}$. Both $\chi$ and
        $V(\chi)$ are given in Planck units.}}
    \label{fig:X_V}
\end{figure*}

We have found an analytic expression for $\chi$ in terms of $\varphi$ from Eqns.~(\ref{eq:ksquared}) and (\ref{eq:varphi-chi}). It is rather complicated however, and we defer the explicit solution to the appendix A.  We plot the result for $\chi$ as a function of $\varphi$ in Fig.~(\ref{fig:X_V}a). The dependence is smooth and linear except at a point where the derivative ${\rm d}\chi/{\rm d}\varphi$ becomes very large (in fact infinite). As a result, the potential $V$ is roughly exponential also as a function of $\chi$ away from the singularity, \emph{i.e.}~automatically of the desired type for quintessence (Eqn.~(\ref{eq:LNquint})). We plot $V(\chi)$ for a representative choice of model parameters in Fig.~(\ref{fig:X_V}b). $V(\chi)$ is indeed exponential for $\chi \lsim -12$.  Moreover it has a minimum that leads to a stabilization of the field, which here corresponds to $\varphiÊ\approx 36$ (or, in the LED context, $M_{\rm b}r \approx 3.5 \times 10^{15}$). Note that the potential minimum is situated close to the ${\rm d}\chi/{\rm d}\varphi$-singularity in $\varphi$. This is so because the $\varphi$-scale is very strongly condensed with respect to $\chi$ near the singularity. We shall return to this and the other features seen in the potential function shortly.

\subsection{The Evolution Equations}

The action (\ref{eq:LEinstein}), written in terms of Einstein frame variables, forms the basis of the cosmology to be studied in the rest of the paper. However, for a homogeneus and isotropic universe, it is the Jordan frame metric $\hat{g}_{\mu \nu}$ which is described by the Friedmann-Robertson-Walker line element:
\begin{equation}
  {\rm d}s^2_{\rm J} \equiv \hat{g}_{\mu \nu} {\rm d}x^{\mu} {\rm d}x^{\nu}
    = {\rm d}t_{\rm J}^2 - R_{\rm J}^2(t_{\rm J}){\rm d}\mathbf{x}^2 \: ,
\end{equation}
where $t_{\rm J}$ and $R_{\rm J}$ are the measured Jordan frame time and scale factor, respectively. Since the relation between the Einstein and Jordan frame metrics is given by $g_{\mu \nu} = F(\varphi)\hat{g}_{\mu \nu}$, the universe is still of Friedmann-Robertson-Walker type in the Einstein frame, with time $t$ and scale factor $R$ given by:
\begin{equation}
  {\rm d}t \equiv \sqrt{F} {\rm d}t_{\rm J} \: , \qquad
  R \equiv \sqrt{F} R_{\rm J} \: .
\end{equation}
Hence, taking a variation with respect to $\chi$ leads to an equation of motion~\footnote{Note that the factor $\sqrt{2}M_{\rm b}$ connected to $\alpha$ in Eqns.~(\ref{eq:EOM}) and (\ref{eq:alpha}) originate in the fact that we have normalized $\chi$ and $V(\chi)$ so that the Lagrangian of the scalar field has the canonical form $\mathcal{L} = \frac{1}{2}(\partial \chi)^2 - V(\chi)$. The coupling strength $\alpha$ is indeed the canonical one, related to the post-Newtonian parameter $\gamma$ via $\gamma - 1 = -2\alpha^2/(1+\alpha^2)$.}
\begin{equation}
  \ddot{\chi} + 3H \dot{\chi} + V'(\chi) +
   \frac{\alpha(\chi)}{\sqrt{2}M_{\rm b}} T^{\mu}_{\mu} = 0\: ,
\label{eq:EOM}
\end{equation}
where $H \equiv \dot{R}/R$ is the Einstein frame Hubble parameter. The explicit form of the derivative of the potential $V'(\chi) \equiv
{\partial V}/{\partial \chi}$ is
\begin{equation}
  V'(\chi) =
  M_{\rm b}^{3} U_0 \frac{1}{(M_{\rm b}r)^8}
\frac{cA - 2aC - 8AC}{A^2 \sqrt{\frac{3}{2} (2A + a)^2 - 2AB}} \: ,
\label{eq:Vprime}
\end{equation}
and the coupling function $\alpha$ is given by
\begin{eqnarray}
  \alpha(\chi) & \equiv & \sqrt{2} M_{\rm b}
        \frac{{\rm d}\log{F^{-1/2}}}{{\rm d}\chi}  \nonumber \\
     & = & - \frac{A + a/2}{\sqrt{\frac{3}{4} (2A + a)^2 - AB}} \: .
\label{eq:alpha}
\end{eqnarray}
These are all exact expressions (at the one-loop) and to evaluate them as a function of $\chi$, one needs the inverse of the transformation (\ref{eq:varphi-chi}), \emph{i.e.} for example $A = 1 + a K^{-1}(\chi )$. Finally, $T^{\mu \nu}$ is the Einstein frame stress-energy tensor of radiation and matter, but since matter is pressureless and $p = \frac{1}{3} \rho$ for radiation, we have simply \mbox{$T^{\mu}_{\mu} = \rho_{\rm m}$}.

The Einstein frame Hubble parameter and the $\alpha$-term couple the evolution of the field $\chi$ to the evolution of the ordinary radiation and matter components. Varying the action (\ref{eq:LEinstein}) with respect to $g_{\mu \nu}$ yields the Friedmann equations for the flat, homogenous and isotropic cosmologies:
\begin{eqnarray}
  3 M_{\rm b}^{2} H^{2} &=& \rho \: ,
\label{eq:FriedmannE1}
\\
  - 2 M_{\rm b}^{2} \dot{H} &=& \rho + p \: ,
\label{eq:FriedmannE2}
\end{eqnarray}
where $\rho$ and $p$ are the total energy density and pressure in the Einstein frame:
\begin{eqnarray}
  \rho &=& \rho_{\rm m}+\rho_{\rm r}+\rho_{\chi} \: \\
  p &=&  p_{\rm r} + p_{\chi} \: .
\end{eqnarray}

At this point it is important to once again remind ourselves of the conformal transformation (\ref{eq:conformal}). In the Jordan frame, the strength of gravity is varying with the field $\varphi$ and the model is characterized by an effective gravitational ``constant'' $G_{\rm eff}$:%
\begin{equation}
  \frac{1}{16\pi G_{\rm eff}} \equiv \frac{M_{\rm b}^2 F(\varphi)}{2} \: .
\end{equation}
However, after conformal transformation to the Einstein frame, the \emph{apparent} strength of gravity $G$ is truly constant:
\begin{equation}
  \frac{1}{16\pi G} \equiv \frac{M_{\rm b}^2}{2} \: .
\end{equation}
The dependence on $\varphi$ instead enters via the the matter Lagrangian, which in the Einstein frame couples to \emph{both} the metric and the scalar field, and shows up as a scaling of masses in the Einstein frame compared to masses in the Jordan frame. This scaling can be deduced from the rest mass of a particle:
$m_{\rm Jordan}^2 \equiv \hat{g}^{\mu \nu} p_{\mu} p_{\nu} =
F(\varphi) g^{\mu \nu} p_{\mu} p_{\nu} \equiv
F(\varphi) m_{\rm Einstein}^2$. That is,
\begin{equation}
 m_{\rm Einstein}
    = \frac{m_{\rm Jordan}}{\sqrt{F(\varphi)}} \: .
\label{eq:mass-scaling}
\end{equation}
Hence, setting $M_{\rm b} = 1$ corresponds to using TeV units in the Jordan frame, but Planck units in the Einstein frame.
% However, note that $M_{\rm b}r = e^{\varphi}$ of course has the same numerical value in both frames since $r$ scales oppositely to $M_{\rm b}$.

Using Eqn.~(\ref{eq:mass-scaling}) together with the usual scale-factor dependence of a 3-volume, one finds
\begin{equation}
  \rho_{\rm m} =
          \rho_{{\rm m}0} \left(\frac{R_0}{R}\right)^3
                 \sqrt{\frac{A_0}{A}} e^{-\varphi + \varphi_0}  \: ,
\end{equation}
where $R$ denotes the Einstein frame scale factor and $R_0$, $\varphi_0$ and $A_0$ refer to the present day values of the quantities.
% This non-trivial evolution of the matter density as well as the coupling term proportional to $\alpha$ in the equation of motion (\ref{eq:EOM}) differentiate our model from simple toy models for quintessence and $k$-essence.
In contrast to pressureless matter, the radiation density $\rho_{\rm r}$ evolves normally ($\rho_{\rm r} \propto R^{-4}$). This can be viewed to arise as a result of the zero rest mass of the photon. Including the contributions from $\chi$ one eventually finds:
\begin{eqnarray}
   \rho &=& \rho_{\rm m0} \left(\frac{R_0}{R}\right)^3
      \sqrt{\frac{A_0}{A}} e^{-\varphi + \varphi_0}
      + \rho_{\rm r0} \left(\frac{R_0}{R}\right)^4 \nonumber \\
      & & {}+ \frac{1}{2} \dot{\chi}^2 + V(\chi) \: , \\
\label{eq:rhofull}
    & & \phantom{-} \nonumber \\
    p &=& p_{\rm r0} \left(\frac{R_0}{R}\right)^4
       + \frac{1}{2} \dot{\chi}^2 - V(\chi) \: .
\label{eq:pfull}
\end{eqnarray}
Equations (\ref{eq:ksquared}--\ref{eq:varphi-chi}) and (\ref{eq:EOM}--\ref{eq:alpha}) together with the Friedmann equations (\ref{eq:FriedmannE1}--\ref{eq:FriedmannE2}) and the explicit expressions for density and pressure (\ref{eq:rhofull}--\ref{eq:pfull}), form a complete set of equations for the evolution of $\chi$ and the Einstein frame scale factor $R$.

% New Section
%%%%%%%%%%%%%%%%%%%%%%%%%%%%%%%%%%%%%%%%%%%%%%%%%%%%%%%%%%%%%%%%%%%%%%%%%%
%
%  --------------------------- Section 4 ------------------------------
%
%%%%%%%%%%%%%%%%%%%%%%%%%%%%%%%%%%%%%%%%%%%%%%%%%%%%%%%%%%%%%%%%%%%%%%%%%%

\section{Stabilization and Convergence of Perturbation Theory}
\label{sec:newstab}

Section \ref{sect:Casimirstab} presented an argument for stabilization of a finite radion in the LED context, based on a particular form of the Casimir potential, Eqn.~(\ref{eq:Udesired}). According to Eqn.~(\ref{eq:largescale}) the stabilization with $e^{\varphi} \equiv M_{\rm b}r \sim 10^{15}$ is consistent with the perturbative scale $|c| \sim \epsilon \sim 1/50$. At first, such a small scale looks perfect for a perturbative treatment of the problem. This is good since the action (\ref{eq:SJordan}) is based on an effective one-loop parametrization of the radiative corrections. There is a fundamental flaw in the above argumentation however, since the actual expansion parameter is not $\epsilon$, but
\begin{equation}
 \varepsilon  \equiv \epsilon \varphi \: .
\end{equation}
Now, reading Eqn.~(\ref{eq:largescale}) the other way around, we see that for $\varphi \equiv \log{M_{\rm b} r} \gg 1$ it sets $\epsilon \simeq 1/\varphi$, or equivalently $\varepsilon \approx 1$!  Hence, the perturbative expansion appears to be necessarily diverging at least in the case of the potential. But since all corrections originate from the same bulk scalar field $\phi$ in the ABRS scenario, they should all be roughly of same order $aÊ\sim  b \sim c \sim 1/50$, which renders the entire one-loop parametrization in Eqns.~(\ref{eq:SJordan}--\ref{eq:C}) useless.  In fact the stabilization mechanism itself would only seem to work qualitatively at best, and could be destroyed by unknown higher order loop and non-perturbative corrections.

It is quite interesting that a similar conclusion also follows from the observational constraints on gravity. Precision tests of General Relativity in the solar system sets a bound on the coupling between the scalar field $\chi$ and matter~\cite{will98}:
\begin{equation}
    \alpha^{2}(\chi_{0}) < 2.2 \times 10^{-4}Ê\equiv \alpha_0^2 \: .
\label{eq:alphaConstraint}
\end{equation}
From Eqn.~(\ref{eq:alpha}) it is then clear that one must have very accurately $A \simeq -a/2$.  From the one-loop expression $A = 1 + a\varphi$ and the fact that $\varphi$ is large by assumption, we then see that $a<0$ and $|a|  \ll 1$, so that a solution only exists if  $a\varphiÊ\simeq -1$. That is, the one-loop contribution must again balance the tree level one to a high precision.

We started out by wanting to view the expansions in Eqns.~(\ref{eq:A}--\ref{eq:C}) as coming from radiative corrections. However, since the radiative corrections need to be large, they cast doubt over the self-consistency of the theory. While the precise forms of the expansions is ultimately irrelevant -- all that matters is that $A$ has a zero at large $\varphi$ (corresponding to a logarithmically large value of $M_{\rm b}r$ in the LED context) -- it is interesting to see that Eqns.~(\ref{eq:A}--\ref{eq:C}) \emph{can} be obtained in the context of a converging perturbation theory. Quite simply, one only needs to assume that the radiative corrections arise not from one, but from several and possibly very many independent, or at most weakly coupled bulk fields $\phi_i$. In such a case the loop corrections from each independent bulk field $\phi_i$ may be small, while their combined one-loop correction can add to a value close to unity. To illustrate the  idea assume that there are $N \gg 1$ completely independent scalar fields $\phi_i$ with equal cubic self-couplings living in the bulk. One would then have a truly small perturbation expansion parameter
\begin{equation}
 \varepsilon_N \equiv \varepsilon/N \sim 1/N
\end{equation}
for each separate bulk field $\phi_i$. In this case, the full perturbation expansions for the coefficients $A, B$ and $C$ appearing in the action (\ref{eq:SJordan}) will be converging. Schematically for example:
\begin{equation}
    A = \sum_{i=1}^N A_i
        = 1 + \# N \varepsilon_N +  \# N \varepsilon_N^2 + \ldots \: ,
\label{eq:Aeqn}
\end{equation}
where $\#$:s represent numbers of order unity. If we now assume that the one-loop contribution is of order one, then the higher order loop corrections are suppressed by powers of $1/N$; for example at the two-loop $N \varepsilon_N^2 \sim \varepsilon/N \sim 1/N$.  Of course it is not necessary to assume that all bulk fields $\phi_i$ have exactly the same couplings. It is enough that the sum of radiative corrections is chosen such that the combined one-loop contribution is of order one at the present value of $\varphi$.  On the other hand, one must prevent large mixing couplings $g_{ijk}$ in order to keep higher order loop corrections small. Indeed, at two-loop in the strongly  couplied limit one would find for example $\delta A^{(2)} \sim N^2\varepsilon_N^2 \sim 1$.

The same qualitative argument also apply to the function $B$. Tuning $A$ close to zero does not imply that $B$ is small at the same time however. Still, the neglected two-loop corrections are small for both quantities. The case of the Casimir potential in the LED context is different; for $U_{\rm C}$ there is no tree level contribution as the whole potential is purely of quantum origin. Its multi-field version corresponding to Eqn.~(\ref{eq:Aeqn}) then looks as follows:
\begin{equation}
    U_{\rm C} = \sum_{i=1}^N U_i
        = N U_N (1 + \# \varepsilon_N +  \# \varepsilon_N^2 + \ldots \:) ,
\end{equation}
After we make the obvious identification $NU_N = U_0 \sim 1$ we observe that the radiative correctios to $U_0$ are truly small: $c \sim \# \varepsilon_N \sim 1/N$. This implies that the original ABRS-type stabilization relying on the corrected Casimir potential is not feasible in this construction.

\subsection{The New Stabilization Mechanism}
\label{sec:NEWstab}

Fortunately, the model does in fact have another stabilization mechanism built in it. Indeed, if the one-loop correction, $\# N \varepsilon_N \equiv a \varphi$ in Eqn.~(\ref{eq:Aeqn}) is negative, then the function $A(\varphi)$ will become zero in the scale $\varphi \simeq 1/|a|$. Since the two-loop corrections to $A$, $B$ and $C$ are still small at this point, they can not alter this conclusion significantly. The new stabilization mechanism now simply follows from the $1/A^2$-scaling of $V_{\rm E}(\varphi)$, which makes it an increasing function of $\varphi$ as $A$ approaches zero, and eventually gives rise to a narrow but infinitely high confining potential barrier at $A=0$.

The potential shown in Fig.~(\ref{fig:X_V}b) actually corresponds to $A=0$ confinement. The minimum seen there has nothing to do with the potential $U$, for which we have in fact used a pure exponential, runaway form with $c=0$. Moreover,  the sharp feature at the top of the potential maximum, the kink, corresponds to the singularity $A(\chi_*)=0$, which would have shown up as an infintely high and narrow spike in $V_{\rm E}(\varphi)$. The fact that the singularity becomes a finite height kink in the $\chi$-picture follows from the highly nontrivial relation between $\chi$ and $\varphi$ (see appendix A). One can understand this when one observes that also $k^2(\varphi)$, Eqn.~(\ref{eq:ksquared}), becomes singular at $A=0$, so that also the kinetic term for $\varphi$ diverges. This has the opposite sign in the action and balances out the potential singularity. (Obviously the model should be qualified as a mixed $k$-essence and quintessence in the singular region.)  Although $V(\chi_*)$ is finite, note that it still is 20 orders of magnitude larger than the value of the potential at the minimum!  Together with Hubble friction this barrier becomes impenetrable, prohibiting the field from reaching values beyond $\chi_*$ and thus giving rise to a robust confinement mechanism for $\varphi$, corresponding to large, finite radii in the LED context.

In summary, we have shown that in a scenario with many weakly coupled independent bulk fields $\phi_i$, the effective theory Eqn.~(\ref{eq:SJordan}) can be reliable even when $A, B \approx 0$.  Stabilization for large values of $\varphi$. corresponding to large radius in the LED context, can not be obtained from the corrected Casimir potential because the scenario predicts $C \approx 1$. However, it can originate from a singularity in the potential at $A=0$.  On the technical side, the fact that the full one-loop contributions to quantities $A$ and $B$ \emph{are} large, implies that one can not simplify any of the expressions that depend on them in the previous sections, by expanding in $a\varphi$ or $b\varphi$ as was done in ref.~\cite{albrecht02_2}. This is of course the reason why we always kept the exact expressions. Finally, it should be stressed that while our quantitative results will depend on the particular form of the one-loop parametrization, the qualitative features would be the same for any function $A$ that crosses zero at a sufficiently large values of $\varphi$.

% New Section
%%%%%%%%%%%%%%%%%%%%%%%%%%%%%%%%%%%%%%%%%%%%%%%%%%%%%%%%%%%%%%%%%%%%%%%%%%
%
%  --------------------------- Section 5 ------------------------------
%
%%%%%%%%%%%%%%%%%%%%%%%%%%%%%%%%%%%%%%%%%%%%%%%%%%%%%%%%%%%%%%%%%%%%%%%%%%

\section{Constraints}
\label{sect:constraints}

The underlying model (\ref{eq:SJordan}) has four unknown parameters: $U_0$, $a$, $b$, and $c$.  Given a full knowledge of the physics leading to the effective theory, these parameters could be calculated exactly in terms of the fundamental coupling constants, but as those details are not known, one has to settle with combinations of phenomenological and theoretical naturalness constraints.  Stabilization arguments, together with the desired large values of $r$ in the LED context, have already settled the overall scale of the parameters $U_0$, $a$, $b$ and $c$. We shall next impose the constraints coming from various observations.

\subsection{Observational Constraints}
\label{sect:obsconstraints}

There are four major observational constraints affecting the model parameters. We have already mentioned the gravity constraint on the scalar field coupling $\alpha$ at present, and we shall return to it again shortly. The other constraints are of astrophysical origin, mainly originating in the CMBR and supernovae observations. As previously stated, it is in the Jordan frame where all observations have their usual interpretation. Hence, to obtain the true constraints on the model coming from for instance the CMBR, one needs to make a full computation of the possible spectra coming from the model and fit them to the observed data. This is however beyond the scope of the present paper.

Instead, we will use a more modest approach based on the fact that the behaviour of our model eventually is very close to that of General Relativity. In such a case, $F(\varphi)$ and $Z(\varphi)$ are essentially constant so that Jordan and Einstein frame quantities do not differ except for overall constant factors. E.g., given that $M_{\rm b}^2 F(\varphi) \sim M_{\rm Pl}^2$, setting $M_{\rm b}$ to one corresponds to using TeV units in the Jordan frame, but Planck units in the Einstein frame. The majority of the constraints concerns the present day value of various quantities. Hence, applying the constraints will essentially only give us the value of our parameters at one specific point (the present) in the evolution of the universe. In practice, we then start the evolution of the universe and shoot for this point. Given that we succeed, we then need to check that the behaviour of the model was close enough to General Realtivity at earlier times. Let us now return to discuss these contraints case by case.

The dark energy component is constrained by it's present amount and by the equation-of-state parameter $w_{\chi}$. The equation-of-state parameter is limited by CMBR and astronomical data to~\cite{wmap}
\begin{equation}
    w_{\chi_0} < -0.78 \: .
\label{eq:wchilimit}
\end{equation}
To be precise, the limit (\ref{eq:wchilimit}) applies for the present value of $w_{\chi}$, and was derived in standard General Relativity with the assumption that both $w_{\chi}$ and $\rho_\chi$ are constants. In the present model both the energy density and the pressure of the quintessence field $\chi$ are dynamical variables and so the bound (\ref{eq:wchilimit}) is not directly applicable. Moreover, the field $\chi$ also couples to regular matter. More general limits on the  function $w_{\chi}(z)$ have been found in the literature~\cite{steenwlimit,otherwlimits}.
% However, parametrizations of the cosmic equation of state in terms of a pure dark energy $w_{\chi}$ do not necessarily lead to an improvement over Eqn.~(\ref{eq:wchilimit}). Indeed, imagine for example that the kinetic energy of a quintessence field is neglible so that $w_{\chi} \approx -1$ to a high accuracy over the range of interest in $z$. The predicitions of the model can still differ significantly from a pure cosmological constant if $V(\chi(z))$ continues to evolve in the scale $\rho_{\rm c}$ (giving rise to a variable $w_{\rm tot}(z)$). This will actually be precisely the case in the examples presented below.
However, we take Eqn.~(\ref{eq:wchilimit}) as an indication that the present value of $w_{\chi}$ should be very close to $-1$. It is convenient to rephrase Eqn.~(\ref{eq:wchilimit}) in terms of present ratio of the kinetic and the potential energy densities of the field $\chi$:
\begin{equation}
    \delta_{\chi_0} \equiv \frac{1 + w_{\chi_0}}{1 - w_{\chi_0}} < 0.12 \: .
\end{equation}

The third observational constraint comes from connecting the model parameters with the observed amount of dark energy at present:
\begin{equation}
   V(\chi_0) = M_{\rm b}^{4} U_0 \frac{C_0}{A_0^{2}} e^{-8\varphi_0}
      \:=\: \frac{\Omega_{\Lambda} \rho_{\rm c}}{1+\delta_{\chi_0}} \: ,
\label{eq:Vconstraint}
\end{equation}
where $\Omega_{\Lambda} \approx 0.73$. The content of Eqn.~(\ref{eq:Vconstraint}) is that a given set of model parameters $U_0$, $a$ and $c$, and the boundary condition for $\delta_{\chi_0}$ will determine the precise value of $\varphi_0$, i.e the value of $\chi$ today. Because $\rho_{\rm c} \sim 10^{-120} M_{\rm b}^{4}$ in the Einstein frame, it is clear that in the LED context, setting $U_0 \sim 1$ imposes roughly the desired scale of $M_{\rm b}r = e^{\varphi} \sim 10^{15}$.

Finally, there is a constraint due to the need of preserving the success of big bang nucleosynthesis: The universe must be radiation dominated already at BBN-time, i.e. when the observed Jordan frame temperature $T_{\rm J} \sim 1$ MeV. This constrains both the kinetic and potential energy densities of the scalar field, which would otherwise make the universe expand too fast and result in too large helium production. This consideration does not lead to additional constraints on model parameters, but rather to a need for tuning the initial conditions for the field. There is a correlation between the two however, because different sets of model parameters need different sets of initial conditions to provide an acceptable cosmological solution. Moreover, it is not \emph{a priori} clear that acceptable solutions do exist at all for arbitrary sets of model parameters. Obviously these questions can ultimately be studied only through numerical analysis.

\subsection{Fine-Tuning in the Model Parameter Space}
\label{sect:fine-tune}

\begin{figure}[!t]  % alpha-constraint
    \begin{center}
	\begin{psfrags}%
	\psfragscanon%
	%
	% text strings:
	\psfrag{s03}[t][t]{\setlength{\tabcolsep}{0pt}
	  \begin{tabular}{c}$b$\end{tabular}}%
	\psfrag{s04}[b][b]{\setlength{\tabcolsep}{0pt}
	  \begin{tabular}{c}$\alpha^2(\chi_0)$\end{tabular}}%
    \psfrag{a01}[r][r]{\small{$a = -0.02755$}}%
    \psfrag{a02}[r][r]{\small{$a = -0.02757$}}%
    \psfrag{a03}[r][r]{\small{$a = -0.02759$}}%
	%
	% xticklabels:
	\psfrag{x01}[t][t]{$-0.1$}%
	\psfrag{x02}[t][t]{$-0.08$}%
	\psfrag{x03}[t][t]{$-0.06$}%
	\psfrag{x04}[t][t]{$-0.04$}%
	\psfrag{x05}[t][t]{$-0.02$}%
	%
	% yticklabels:
	\psfrag{v01}[r][r]{$0$}%
	\psfrag{v02}[r][r]{$2$}%
	\psfrag{v03}[r][r]{$4$}%
	\psfrag{v04}[r][r]{$6$}%
	\psfrag{v05}[r][r]{$8$}%
	\psfrag{v06}[r][r]{$10$}%
	\psfrag{ypower1}[Bl][Bl]{$\times 10^{-4}$}%
	%
    % Figure:
    \includegraphics[width=7.5cm]{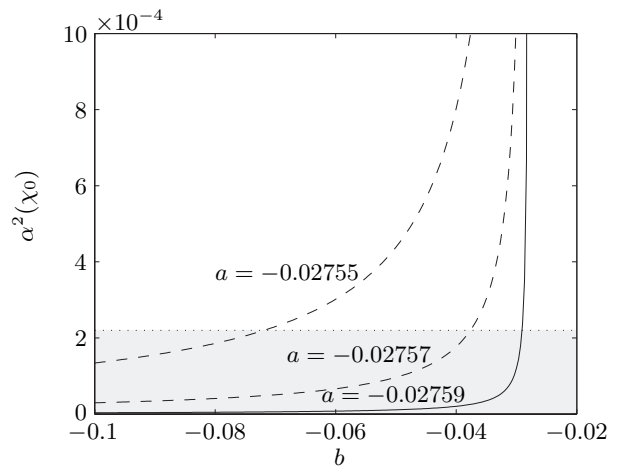}%
    \end{psfrags}%
    \end{center}
    \caption{\small{Shown is the present value of the square of the 
        coupling $\alpha$ as a function of the parameter $b$ for three
        different values of the parameter $a$. The other model parameters
        are set to the values given in Eqn.~(\ref{eq:refset}).
        The $\alpha$-bound, Eqn.~(\ref{eq:alphaConstraint}), allows the
        region below the dotted line.}}
        \label{fig:alphaConstraint}
\end{figure}
\begin{figure*}[!t] % k-squared as a function Mbr
    \begin{center}
    \begin{psfrags}%
    \psfragscanon%
    %
    % text strings:
    \psfrag{f01}[t][t]{\setlength{\tabcolsep}{0pt}
      \begin{tabular}{c}Fig. (a)\end{tabular}}%
    \psfrag{f02}[t][t]{\setlength{\tabcolsep}{0pt}
      \begin{tabular}{c}Fig. (b)\end{tabular}}%
    \psfrag{s03}[t][t]{\setlength{\tabcolsep}{0pt}
      \begin{tabular}{c}$\varphi/\log{10}$\end{tabular}}%
    \psfrag{s04}[b][b]{\setlength{\tabcolsep}{0pt}
      \begin{tabular}{c}$k^2(\varphi)$\end{tabular}}%
    \psfrag{s10}[l][l]{\small{$a = -0.02759$}}%
    \psfrag{s11}[l][l]{\small{$b = -0.034$}}%
    \psfrag{s12}[l][l]{\small{$b = -0.031$}}%
    \psfrag{s13}[l][l]{\small{$a = -0.037$}}%
    \psfrag{s14}[l][l]{\small{$b = -0.036$}}%
    \psfrag{s15}[l][l]{\small{$a = -0.031$}}%
    \psfrag{s16}[l][l]{\small{$b = -0.026$}}%
    %
    % xticklabels:
    \psfrag{x01}[t][t]{$10$}%
    \psfrag{x02}[t][t]{$12$}%
    \psfrag{x03}[t][t]{$14$}%
    \psfrag{x04}[t][t]{$16$}%
    \psfrag{x05}[t][t]{$18$}%
    \psfrag{x06}[t][t]{$20$}%
    %
    % yticklabels:
    \psfrag{v01}[r][r]{$-10$}%
    \psfrag{v02}[r][r]{$0$}%
    \psfrag{v03}[r][r]{$10$}%
    \psfrag{v04}[r][r]{$20$}%
    \psfrag{v05}[r][r]{$30$}%
    %
    % Figure:
    \includegraphics[width=14.5cm]{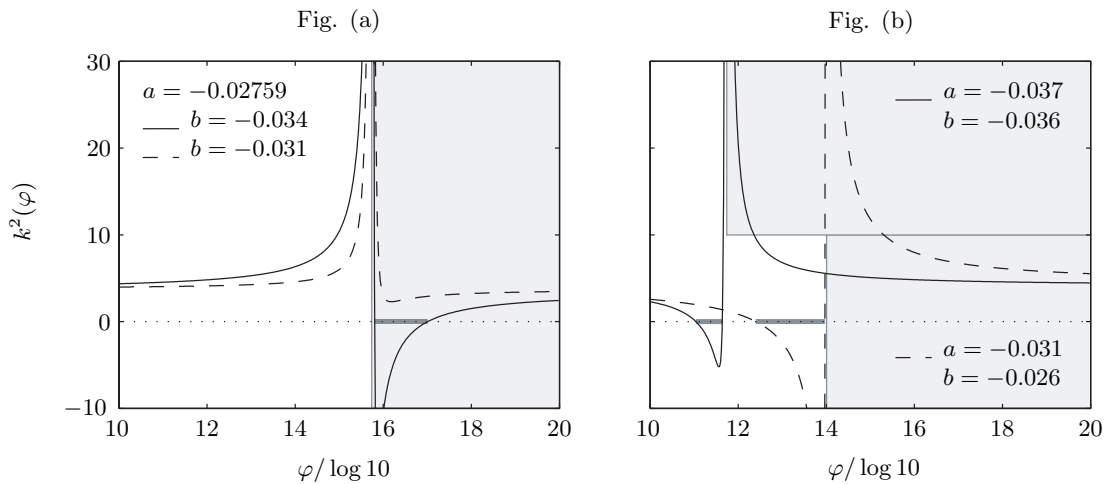}%
    \end{psfrags}%
    \end{center}
    \caption{\small{$k^2(\varphi)$ as a function of
    	$\varphi/\log{10}$ ($= \log_{10} M_{\rm b}r$)  for different values of
    	the parameters $a$ and $b$.  Thick gray lines highlight the values of
    	$\varphi$ corresponding to a phantom phase. Shaded regions correspond
    	to values of $\varphi$ where gravitons would carry negative energy in
    	the Jordan frame.}}
    \label{fig:ksquared}
\end{figure*}
Let us now return to the gravity constraint Eqn.~(\ref{eq:alphaConstraint}) ($\alpha$-bound for short). While the form of the potential automatically adjusts $A$ close to zero, it is not immiediately clear whether this tuning is sufficient to satisfy the $\alpha$-bound automatically. Comparing Eqns.~(\ref{eq:Vprime}) and (\ref{eq:alpha}) we see that there is in fact some tension between fulfilling the $\alpha$-bound and reaching the potential minimum where $V'(\chi) = 0$. Indeed, for $c = 0$ the latter corresponds to $A = - a/4$, whereas the former goes to zero at $A = - a/2$. Whether this difference can be accommodated obviously depends on how strict the $\alpha$-bound really is.  From the parametric dependence of $\alpha = \alpha(\chi, a, b)$ one sees that the $\alpha$-bound leads to correlations between the model parameters, and so does the need to adjust $V = V(\chi,U_0,a,c)$ to the present cosmological constant as well. In the end one is forced to some cross-correlated fine-tuning with the model parameters.  Instead of trying to map out the allowed region in the parameter space, we estimate the degree of the fine-tuning that is needed by fixing all other parameters to some natural values and then checking how much freedom is left in varying $a$.  Observe that this does not mean that $a$ is \emph{a priori} the parameter to be fine tuned; we could as well set $a$ rather freely, and then fine-tune either $U_0$, or even $\chi_0$, i.e. the radius of the extra dimensions today $r_0$ in the LED context.

Let us assume that $\chi_0$ is fixed. For definiteness we assume that the potential parameters and $\delta_{\chi_0}$ take on the values
\begin{equation}
    U_0 = 2.60 \:, \quad c = 0 \quad \textrm{and} \quad
    \delta_{\chi_0} = 10^{-2} \: .
\label{eq:refset}
\end{equation}
For a fixed $\chi_0$ the $\alpha$-bound implies that $A_0+a/2$ has to be
fine-tuned to zero to an accuracy
\begin{equation}
    |A_0 + a/2| \le \sqrt{\alpha_{\rm 0}^2 |aB_0| / 2} \: .
\label{eq:fine-tune}
\end{equation}
Because of the smallness of $\alpha_{\rm 0}^2$ this is always a rather strong constraint, and it becomes particularily severe when $|B_0|$ is small. This dependence is clearly visible in Fig.~(\ref{fig:alphaConstraint}), where we plot $\alpha$ as a function of $b$ for some representative values of $a$. For the values chosen above, $|B_0|$ goes to zero at $b_* \approx -0.0280$~\footnote{One can find through Eqns.~(\ref{eq:Vconstraint}) and (\ref{eq:refset}), together with the \emph{a posteriori} knowledge that $a$ is almost fixed, that $\varphi_0 \approx 35.8$ and then $b_* = -1/\varphi_0 \approx -0.0280$.} and the closer $b$ gets to $b_*$, the more accurately $a$ must be fine tuned to a value $a \to b_*/(1 - b_*/2) \approx -0.0276$. If we now impose the perturbative argument that $|b|$ is of the same order as $|a|$, it implies that $a$ has to be fine-tuned to roughly one part in ten to hundred thousand, which corresponds to a relative accuracy of somewhat less than one per cent. This gives an estimate of the amount of fine-tuning that is needed when adjusting different acceptable model parameter sets.  In what follows we shall fix $a$ to the value that allows the widest interval in $b$ in Fig.~(\ref{fig:alphaConstraint}), \emph{i.e.}
\begin{equation}
        a = -0.02759 \: \qquad {\rm and}Ê\qquad b < -0.0300 \: .
\label{eq:aandb}
\end{equation}
Eqns.~(\ref{eq:refset}) and (\ref{eq:aandb}) define the allowed parameter space for our quintessence-like scenarios with the model action (\ref{eq:LEinstein}). Changing $U_0$ and $c$ only moves solutions to a slightly different band in $a$ and $b$ space. To be specific, we will choose $b= -0.034$ as our reference value in what follows.

Finally note that the expression on the right-hand side of Eqn.~(\ref{eq:fine-tune}) in general is smaller than $|a|/4 \approx 0.007$, the value of $|A|$ corresponding to the minimum of $V(\chi)$. Indeed, for our reference value of $b$ we have $\sqrt{\alpha_{\rm 0}^2 |aB_0|} \approx 0.001$. As a result the cosmological evolution must be such that the field is $\chi$ actually \emph{not} in its potential minimum at present. Moreover, all our solutions should show some late time evolution and adjustment of $\alpha$ close to the present time.  One can lessen this effect by choosing a large value for $|B_0|$, but apart from such additional tuning the late evolution should be a generic prediction of the model.

\subsection{Phantom Parametrizations}

For our reference set, Eqns.~(\ref{eq:refset}) and (\ref{eq:aandb}) with $b = -0.034$, the model stays non-phantom for all values of $\varphi$ with $F(\varphi) > 0$. As seen in Fig.~(\ref{fig:ksquared}a), the field $\varphi$ becomes phantom over a finite interval in the region where gravitons would carry negative energy in the Jordan frame, but since the field evolution never enters the this region, such cosmologies cannot be realized from natural initial conditions. However, it is also possible to arrange the model parameters such that $\varphi$ becomes a phantom field already before the $A=0$ singularity. We show examples of such models in Fig.~(\ref{fig:ksquared}b). It is quite interesting that these phantom-like structures were found using perfectly natural values for the model parameters. Nevertheless, it is not clear to us whether one can arrange a phantom phase into the past, present or the future of \emph{our} universe such that it would be compatible with all the observational constraints. At least the examples that we have studied connected to large extra dimensions, including the models shown in Fig.~(\ref{fig:ksquared}b), almost certainly are ruled out by gravity constraints. However, if one indeed takes the generic scalar-tensor theory of Eqn.~(\ref{eq:S}) as the starting point, several restrictions such as the millimeter scale for large extra dimensions no longer apply, and one could obtain a richer structure of possibly acceptable mixed phantom-quintessence-$k$-essence solutions.

A promising study addressing these questions has been done by Perivolaropoulos~\cite{perivol} who show that, in contrast to the case of a minimally coupled scalar field~\cite{vikman}, it is indeed possible to reproduce an eventual phantom transition within scalar-tensor theories. Perivolaropoulos employs a link between the equation of state of dark energy and the precision tests of General Relativity in the solar system. In this scheme, one redefines $\varphi$ in the Jordan frame so that the field has a canonical kinetic term (apart from the sign). Supernovae data and the GR-constraint are then used to put bounds on the redshift $z$ dependence of $\varphi$, the potential $U$, and the function $F$ multiplying $R_{\hat{g}}$ in the Jordan frame action. By assuming a particular form of $\varphi(z)$ it is then possible to reconstruct $U(\varphi)$ and $F(\varphi)$. We leave such explorations for upcoming studies. The conditions for obtaining a dark energy equation of state less than $-1$ has also been explored in ref.~\cite{martin}.
\begin{table}[!t]
    \begin{center}
    \begin{tabular}{|c|c|c|}
        \hline
            \small{\textbf{Variable}}
            & \small{\textbf{Starting Value}}
            & \small{\textbf{Present Value}} \\
        \hline
        \hline
            $\log_{10}{R_{\rm J}}$
            & $-11.5$
            & $0.00$ \\
        \hline
            $\log_{10}{R}$
            & $-12.0$
            & $0.00$ \\
        \hline
            $\chi$
            & $-31.6$
            & $-24.1$ \\
        \hline
            $\partial \chi / \partial t$
            & $2.0 \times 10^{-37}$
            & $2.57 \times 10^{-61}$ \\
        \hline
            $\varphi$
            & $33.7$
            & $35.8$ \\
        \hline
            $\rho_{\chi}$
            & $2.0 \times 10^{-74}$
            & $8.42 \times 10^{-121}$ \\
        \hline
            $\rho_{\rm m}$
            & $1.1 \times 10^{-84}$
            & $3.16 \times 10^{-121}$ \\
        \hline
            $\rho_{\rm r}$
            & $2.9 \times 10^{-76}$
            & $2.90 \times 10^{-124}$ \\
        \hline
            $\rho_{\rm tot}$ & -- & $1.16 \times 10^{-120}$ \\
        \hline
            $\Omega_{\chi}$ & -- & $0.727$ \\
        \hline
            $\Omega_{\rm m}$ & -- & $0.272$ \\
        \hline
            $\Omega_{\rm r}$ & -- & $2.50 \times 10^{-4}$ \\
        \hline
            $w_{\chi}$ & -- & $-0.922$ \\
        \hline
    \end{tabular}
    \end{center}
    \caption{\small{Initial conditions/starting values and the resulting
        values at present, given in Planck units. Note that
        $\rho_{\rm tot} = \rho_{\rm c}$.}}
    \label{table:startPresent}
\end{table}

\subsection{Other Constraints}

Some of the constraints discussed so far were generic to any effective quintessence model. Others, such as the need for $\varphi$ to have the same value at BBN-scale as at present, were particular to the present model having its origin in scalar-tensor theory. In the LED context, there is yet another type of constraint particular to these models that we need to discuss. Namely, the radion degree of freedom that we have seen to give rise to quintessence field, comes associated with an infinite tower of Kaluza-Klein excitations on the brane. The problem with the KK-modes is that if the universe enters the radiation dominance at too high temperature $T_*$, too many KK modes might be excited through their couplings with ordinary particles~\cite{Hamedetal,Hall,steen}. However, it is quite difficult to state precisely how these considerations limit the model parameters, because they depend on introducing an entire new structure to the model, which has not yet been explored in detail. Indeed, the problem of KK-modes might be associated and solved by the unknown higher-dimensional physics, in which case even no additional constraints might arise~\cite{albrecht02_2}. Even in the more conservative view, which assumes the possiblity of exciting the KK-modes at the reheating scale $T_*$, bounds depend on the details of the temperature history on the brane. They can be quite stringent however; for example, Hannestad~\cite{steen} reports that if the universe enters radiation dominace instantly, then the fundamental mass scale (having fixed $M_{\rm b}r$) is bounded by $M_{\rm b} \gsim 8 + 10(T_*/{\rm MeV})$ TeV.  In principle we can always accommodate such a constraint in our model. This would of course to some degree undermine the particle physics motivated attraction to it, which originates from connecting the fundamental gravity and electroweak scales to evade the heirarchy problem. From a cosmology point of view, the main problem is however that an increase in $M_{\rm b}$ must be compensated by a decrease in $r$, which would eventually spoil the naturalness of the Casimir potential as the source of dark energy.  The situation is not extreme however; taking a reheating temperature $T_* = 1$ MeV, gives $r \lsim (0.02 {\rm eV})^{-1}$, which would still imply a reasonably modest fine-tuning of order $U_0 \sim 10^{-5}$.

% New Section
%%%%%%%%%%%%%%%%%%%%%%%%%%%%%%%%%%%%%%%%%%%%%%%%%%%%%%%%%%%%%%%%%%%%%%%%%%
%
%  --------------------------- Section 6 ------------------------------
%
%%%%%%%%%%%%%%%%%%%%%%%%%%%%%%%%%%%%%%%%%%%%%%%%%%%%%%%%%%%%%%%%%%%%%%%%%%

\section{A Realistic cosmology}
\label{sect:realistCosm}

\begin{figure}[!t]  % Jordan frame scale factor
    \begin{center}
	\begin{psfrags}%
	\psfragscanon%
	%
	% text strings:
	\psfrag{s03}[t][t]{\setlength{\tabcolsep}{0pt}%
		\begin{tabular}{c}$\log_{10} R$\end{tabular}}%
	\psfrag{s04}[b][b]{\setlength{\tabcolsep}{0pt}%
		\begin{tabular}{c}$\Delta R \:, \; \Delta t$\end{tabular}}%
	%
	% xticklabels:
	\psfrag{x01}[t][t]{$-12$}%
	\psfrag{x02}[t][t]{$-10$}%
	\psfrag{x03}[t][t]{$-8$}%
	\psfrag{x04}[t][t]{$-6$}%
	\psfrag{x05}[t][t]{$-4$}%
	\psfrag{x06}[t][t]{$-2$}%
	\psfrag{x07}[t][t]{$0$}%
	%
	% yticklabels:
	\psfrag{v01}[r][r]{$-1$}%
	\psfrag{v02}[r][r]{$-0.5$}%
	\psfrag{v03}[r][r]{$0$}%
	\psfrag{v04}[r][r]{$0.5$}%
	\psfrag{v05}[r][r]{$1$}%
	%
    % Figure:
    \includegraphics[width=7.5cm]{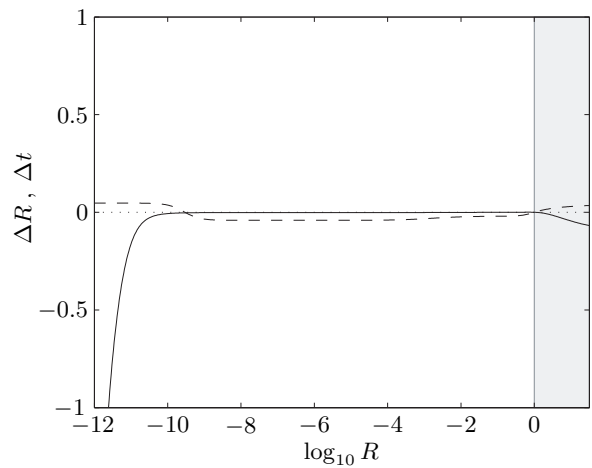}%
    \end{psfrags}%
    \end{center}
    \caption{\small{The relative difference between the Einstein and the Jordan
    	frame scale factors $\Delta R \equiv (R - R_{\rm J})/R$ (solid) and
    	times $\Delta t \equiv (t - t_{\rm J})/t$ (dashed), where we have set
    	$R = R_{\rm J} = 1$ and $t = t_{\rm J}$ at present.}}
    \label{fig:RtJordan}
\end{figure}
In the previous section we showed that setting the fundamental energy scale to TeV, or equivalently, the stabilized extra dimensions to millimeter scale in the LED context, combined with theoretical and observational constraints practically fixes the fundamental parameters in the model. This means that the form of the quintessence potential $V(\chi)$ is almost set as well, not leaving much extra space for adjustments to get an acceptable cosmological evolution. Of course, a lot has already been achieved in the course of fixing the parameters; built in is the feature that the potential has a minimum, corresponding to millimeter scale in the LED context, such that $V(\chi)$ would give the right amount of dark energy today. Also, the gravity constraints and constraints on the equation-of-state parameter are almost automatically satisfied if the field sits near the potential minimum. However, we still have to show that the model indeed does behave very similar to General Relativity and that it can support a cosmological evolution consistent with all constraints, leading to the desired densities at present.

\begin{figure}[!t]  % Energy densities
    \begin{center}
    \begin{psfrags}%
    \psfragscanon%
    %
    % text strings:
    \psfrag{s03}[t][t]{\setlength{\tabcolsep}{0pt}
       \begin{tabular}{c}$\log_{10}{R}$\end{tabular}}%
    \psfrag{s04}[b][b]{\setlength{\tabcolsep}{0pt}
    \begin{tabular}{c}$\log_{10}{\rho}$\end{tabular}}%
	\psfrag{A}[t][t]{\large{$\downarrow$}}%
	\psfrag{t01}[l][l]{\small{$T_{\rm J} = 1$ MeV}}%
    %
    % xticklabels:
    \psfrag{x01}[t][t]{$-12$}%
    \psfrag{x02}[t][t]{$-10$}%
    \psfrag{x03}[t][t]{$-8$}%
    \psfrag{x04}[t][t]{$-6$}%
    \psfrag{x05}[t][t]{$-4$}%
    \psfrag{x06}[t][t]{$-2$}%
    \psfrag{x07}[t][t]{$0$}%
    %
    % yticklabels:
    \psfrag{v01}[r][r]{$-130$}%
    \psfrag{v02}[r][r]{$-120$}%
    \psfrag{v03}[r][r]{$-110$}%
    \psfrag{v04}[r][r]{$-100$}%
    \psfrag{v05}[r][r]{$-90$}%
    \psfrag{v06}[r][r]{$-80$}%
    \psfrag{v07}[r][r]{$-70$}%
    %
    % SUBPLOT
        %
        % text strings:
        \psfrag{s13}[t][t]{\setlength{\tabcolsep}{0pt}
            \begin{tabular}{c}\small{$\log_{10} R$}\end{tabular}}%
        \psfrag{s14}[b][b]{\setlength{\tabcolsep}{0pt}
            \begin{tabular}{c}\small{$\rho$}\end{tabular}}%
        %   
        % xticklabels:
        \psfrag{x11}[t][t]{\small{$-1$}}%
        \psfrag{x12}[t][t]{\small{$-0.5$}}%
        \psfrag{x13}[t][t]{\small{$0$}}%
        %
        % yticklabels:
        \psfrag{v11}[r][r]{\small{$8$}}%
        \psfrag{v12}[r][r]{\small{$9$}}%
        \psfrag{v13}[r][r]{\small{$10$}}%
        \psfrag{ypower2}[Bl][Bl]{\small{$\times 10^{-121}$}}%
    %
    % Figure:
    \includegraphics[width=7.5cm]{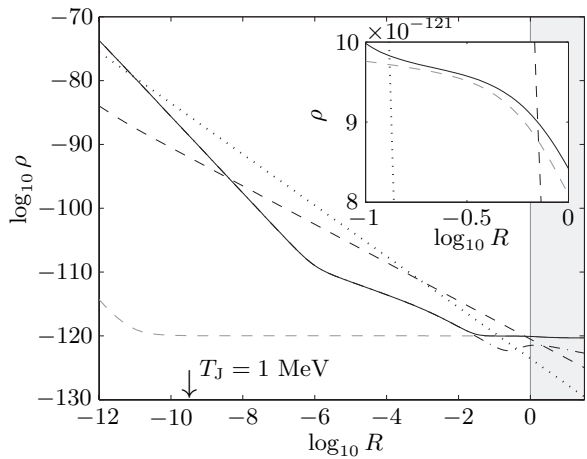}%
    \end{psfrags}%
    \end{center}
    \caption{\small{Different components of the energy density (in Planck units)
        as a function of the Einstein frame scale factor $R$. The various curves
        represent matter (dashed black), radiation (dotted) and scalar energy
        density (solid). The dashed grey and dot-dashed curves shows the
        potential and the kinetic part of the scalar energy density,
        respectively.}}
    \label{fig:rho}
\end{figure}
Let us stress that we actually do not know the full theory at the fundamental TeV-scale; we have to our disposal only an effective theory that is valid at much lower energy scales. In fact, the thermal history beyond the entrance to radiation dominance is subject to unknown physics such as Kaluza-Klein excitations in the LED context and the constraints arising thereof~\cite{Hamedetal,Hall,steen}. We will therefore start off the evolution of our model universe at a temperature well below the TeV-scale, $T_{\rm J} \ll 1$ TeV, but still above the big bang nucleosynthesis temperature of $T_{\rm J} \sim 1$ MeV. In practice we are using the Einstein frame scale factor $R$ as an integration variable and we set $R_{\rm now} \equiv R^{\rm J}_{\rm now} \equiv 1$. We know that $R^{\rm J}_{\rm BBN} \simeq 10^{-10}-10^{-9}$, and $R^{\rm J}_{\rm TeV} \simeq 10^{-16}$, but it is not possible to determine the corresponding values of the Einstein frame scale factor until we know the actual evolution. However, since we have assumed that the model should behave very close to General Relativity, the Einstein frame scale factor should not differ significantly from the Jordan frame scale factor. Having chosen a suitable starting point $R_{\rm start}$, we integrate the evolution equations until present with variable initial conditions for $\chi$ and $\dot \chi$ until (if) an acceptable solution is found. In this scheme, the first and most important test is to see if our evolution indeed behaves like General Relativity after nucleosynthesis. Having established that, we can then check if other quantities such as energy densities fulfils the desired constraints.

\begin{figure}[!t]  % Equation-of-state parameters
    \begin{center}
    \begin{psfrags}%
    \psfragscanon%
    %
    % text strings:
    \psfrag{s03}[t][t]{\setlength{\tabcolsep}{0pt}
       \begin{tabular}{c}$\log_{10}{R}$\end{tabular}}%
    \psfrag{s04}[b][b]{\setlength{\tabcolsep}{0pt}
       \begin{tabular}{c}$w$\end{tabular}}%
    %
    % xticklabels:
    \psfrag{x01}[t][t]{$-12$}%
    \psfrag{x02}[t][t]{$-10$}%
    \psfrag{x03}[t][t]{$-8$}%
    \psfrag{x04}[t][t]{$-6$}%
    \psfrag{x05}[t][t]{$-4$}%
    \psfrag{x06}[t][t]{$-2$}%
    \psfrag{x07}[t][t]{$0$}%
    %
    % yticklabels:
    \psfrag{v01}[r][r]{$-1$}%
    \psfrag{v02}[r][r]{$-0.5$}%
    \psfrag{v03}[r][r]{$0$}%
    \psfrag{v04}[r][r]{$0.5$}%
    \psfrag{v05}[r][r]{$1$}%
    %
    % Figure:
    \includegraphics[width=7.5cm]{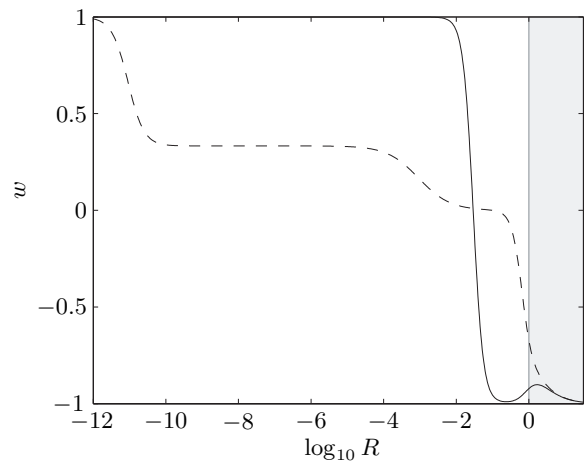}%
    \end{psfrags}%
    \end{center}
    \caption{\small{The equation-of-state parameters $w_{\chi}$ (solid)
        and $w_{\rm tot}$ (dashed) as functions of the Einstein frame
        scale factor $R$.}}
    \label{fig:w}
\end{figure}
We now turn to the qualitative features to be met by an acceptable solution. First one has to make sure that the model does not spoil the BBN explanation for the observed light element abundances.  As was explained in the previous section, this is ensured if $\varphi$ was already very close to its present value during BBN. This turns out to be a tough constraint, which basically forces us to set the field relatively close to its potential minimum at $R_{\rm start}$. Fixing BBN combined with the constraints built into our model potential, go a long way towards fixing the correct cosmological model today.  What remains is essentially the issue of getting the quantitative values of present energy densities  $\Omega_{\chi} \approx 0.73$ and $\Omega_{\rm m} \approx0.27$ to coincide with $\rho_{\rm tot} \approx \rho_{\rm c}$. This of course is just the familiar ``why now" problem, which in general requires some fine-tuning of initial conditions in generic quintessence models. Here the "why now" problem is essentially transformed to the assumed TeV scale of $M_{\rm b}$, corresponding to millimeter size extra dimensions in the LED context.

\begin{figure*}[!t] % Mbr and alpha
    \begin{center}
    \begin{psfrags}%
    \psfragscanon%
    %
    % text strings Fig (a):
    \psfrag{f01}[t][t]{\setlength{\tabcolsep}{0pt}
      \begin{tabular}{c}Fig. (a)\end{tabular}}%
    \psfrag{s05}[t][t]{\setlength{\tabcolsep}{0pt}
       \begin{tabular}{c}$\log_{10}{R}$\end{tabular}}%
    \psfrag{s06}[b][b]{\setlength{\tabcolsep}{0pt}
       \begin{tabular}{c}$\varphi/\log{10}$\end{tabular}}%
    \psfrag{s07}[t][b]{\setlength{\tabcolsep}{0pt}
       \begin{tabular}{c}$k^2(\varphi)$\end{tabular}}%
    %
    % text strings Fig (b):
    \psfrag{f02}[t][t]{\setlength{\tabcolsep}{0pt}
      \begin{tabular}{c}Fig. (b)\end{tabular}}%
    \psfrag{s03}[t][t]{\setlength{\tabcolsep}{0pt}
        \begin{tabular}{c}$\log_{10} R$\end{tabular}}%
    \psfrag{s04}[b][b]{\setlength{\tabcolsep}{0pt}
        \begin{tabular}{c}$\alpha$\end{tabular}}%
    %
    % xticklabels:
    \psfrag{x01}[t][t]{$-12$}%
    \psfrag{x02}[t][t]{$-9$}%
    \psfrag{x03}[t][t]{$-6$}%
    \psfrag{x04}[t][t]{$-3$}%
    \psfrag{x05}[t][t]{$0$}%
    %
    % Mbr yticklabels:
    \psfrag{r01}[r][r]{$14$}%
    \psfrag{r02}[r][r]{$14.5$}%
    \psfrag{r03}[r][r]{$15$}%
    \psfrag{r04}[r][r]{$15.5$}%
    \psfrag{r05}[r][r]{$16$}%
    %
    % k^2 yticklabels:
	\psfrag{k01}[r][r]{$0$}%
	\psfrag{k02}[r][r]{$10$}%
	\psfrag{k03}[r][r]{$20$}%
	\psfrag{k04}[r][r]{$30$}%
	\psfrag{k05}[r][r]{$40$}%
	\psfrag{k06}[r][r]{$50$}%
	\psfrag{k07}[r][r]{$60$}%
	\psfrag{k08}[r][r]{$70$}%
    %
    % alpha yticklabels:
    \psfrag{v01}[r][r]{$-0.4$}%
    \psfrag{v02}[r][r]{$-0.2$}%
    \psfrag{v03}[r][r]{$0$}%
    \psfrag{v04}[r][r]{$0.2$}%
    %
    % SUBPLOT
		%
		% text strings:
		\psfrag{s08}[b][b]{\setlength{\tabcolsep}{0pt}
			\begin{tabular}{c}$\alpha$\end{tabular}}%
		%
		% xticklabels:
		\psfrag{x11}[t][t]{$-1$}%
		\psfrag{x12}[t][t]{$-0.5$}%
		\psfrag{x13}[t][t]{$0$}%
		%
		% yticklabels:
		\psfrag{v11}[r][r]{$-0.02$}%
		\psfrag{v12}[r][r]{$0$}%
		\psfrag{v13}[r][r]{$0.02$}%
		%
	%
    % Figure:
    \includegraphics[width=15cm]{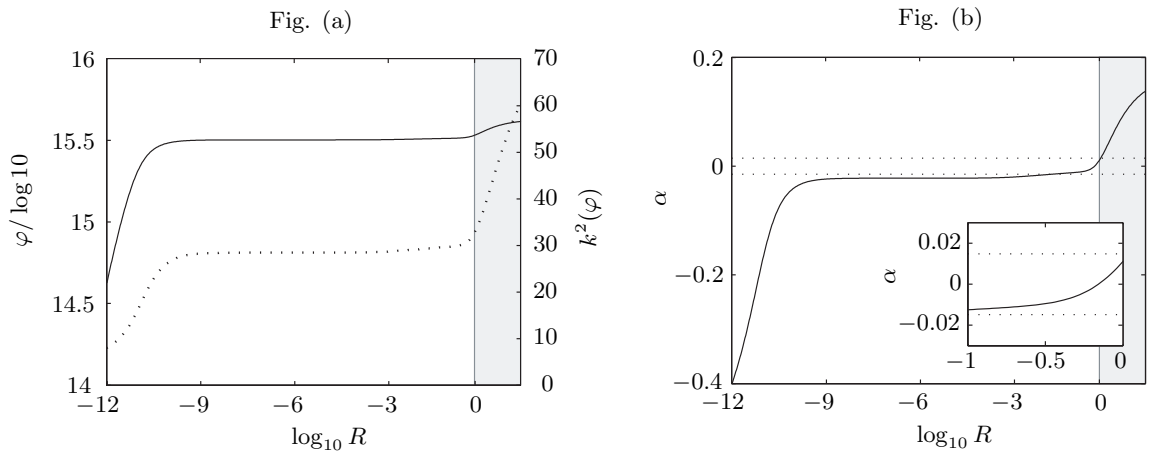}%
    \end{psfrags}%
    \end{center}
    \caption{\small{(a) Shown is $\varphi/\log{10}$
        ($= \log_{10} M_{\rm b}r$) (solid line) and the
        $k^2(\varphi)$-factor defined in Eqn.~(\ref{eq:ksquared})
        (dotted line) as a function of the Einstein frame scale factor $R$.
        (b) The coupling $\alpha$ as a function of the Einstein frame scale
        factor $R$.}}
    \label{fig:Mbr_alpha}
\end{figure*}
Below we present an example of a realistic cosmology corresponding to our basis set of model parameters; $U_0 = 2.60$, $a = -0.02759$, $b = -0.034$, and $c = 0$. We have started the evolution at $R_{\rm start} = 10^{-12}$, which roughly corresponds to the scale $T^{\rm J}_{\rm start} \sim 100$ MeV. The corresponding initial conditions/present values of $\chi$, $\dot{\chi}$ etc. are given in Table~\ref{table:startPresent}, and the cosmological evolution of $R_{\rm J}$ and $t_{\rm J}$, energy densities $\rho_i$, the equation-of-state parameter $w_{\chi}$, the field $\varphi$ and the coupling $\alpha$ are shown in Figs.~(\ref{fig:RtJordan}), (\ref{fig:rho}), (\ref{fig:w}), (\ref{fig:Mbr_alpha}a), and (\ref{fig:Mbr_alpha}b), respectively. Note that we have extended the evolution a bit further than present, with grey areas representing the future.

First of all, we see from Fig.~(\ref{fig:RtJordan}) that, except at very early times, the relative differences between the Einstein and the Jordan frame times and scale factors are neglible, so that the frames are descriptively equivalent. That is, the rate of Einstein frame clocks and size of Einstein frame rods essentially do not differ from the standard Jordan frame clocks and rods except for a constant scaling. Hence, our model does in practice behave like General Relativity and our application of the constraints in section~\ref{sect:obsconstraints} remain valid. This also tells us that $\varphi$ has remained very close to constant, so that the strength of gravity had the same value at BBN as at present. Let us now proceed with discussing the evolution in the Einstein frame in detail.

Due to the relatively large starting value of $\varphi$, the potential is neglible compared to the kinetic energy and the scalar field $\chi$ enters radiation domination in a kinetic-dominated roll. During early stages, the kinetic energy scales as $\dot{\chi}^2 \propto  R^{-6}$, and it dominates the energy density for $\log_{10}{R} \lsim -11$. This scaling follows from Eqn.~(\ref{eq:EOM}) when the potential and the $\alpha$-term are neglected and the evolution of the scalar field is dominated by the Hubble friction $3H\dot{\chi}$. During this \emph{kination} phase~\cite{joyce} the field $\chi$ is rolling down its potential and $\varphi$, corresponding to the radius in the LED context, is growing rapidly. The scalar field dominance ends and the universe enters radiation domination right before BBN. At this time $\chi$ settles close to the minimum of the potential, and correspondingly $\varphi$ reaches a plateau as shown in Fig.~(\ref{fig:Mbr_alpha}a). The energy of the scalar field continues to fall rapidly, until kination is halted by the $\alpha$-term. The $\alpha$-term causes $\dot \chi^2$ to track matter density until $\log_{10}{R} \approx -1.5$, corresponding to and observed redshift $z_{\rm J} \approx 30$~\footnote{The observed redshift $z_{\rm J}$ is naturally defined via $z_{\rm J} + 1 \equiv R_{\rm J0}/R_{\rm J}$, where $R_{\rm J}$ is the Jordan frame scale factor.}, where the kinetic energy falls below the potential, and the field $\chi$ starts acting like a cosmological constant.

Although $V(\chi)$ now dominates the total energy density, it is not yet a constant (see the linear scale plot inserted in Fig.~(\ref{fig:rho})). The cause for this final 10\% decrease between $z_{\rm J} \approx 0.5$ and present can be tracked to the tension between the $\alpha$-bound and the position of the potential minimum for our model parameters, discussed in the last paragraph of section \ref{sect:fine-tune}. Indeed, for a long time, the $\alpha$-term acts like an additional effective potential in the equation of motion (\ref{eq:EOM}) and confines the field to the proximity of the zero of the $\alpha$-function at $A \sim -a/2$.~\footnote{This behaviour of the $\alpha$-term has been discovered elsewhere in the context of chameleon models~\cite{chameleon}.} As the matter density becomes very small, the $\alpha$-term eventually becomes small in comparision to the potential gradient, and the field can finally roll down to its true minimum. This phase can also be seen as a slight increase in $\varphi$ in Fig.~(\ref{fig:Mbr_alpha}a), as well as in a little more dramatic change in $k^2(\varphi)$  and finally the coupling $\alpha$ itself shown in Fig.~(\ref{fig:Mbr_alpha}b).
% For the particular solution we have displayed in detail, the $\alpha$-bound is in fact barely satisfied at present. Our discussion in section \ref{sect:fine-tune} suggests that we can do better by using a somewhat larger magnitude for $b$, since this allows for more flexibility in $a$ and hence makes it easier to satisfy the $\alpha$-bound. We can see that this line of reasoning works from Fig~(\ref{fig:Mbr_alpha}b), which also shows an evolution in $\alpha$ corresponding to $b = -0.034$, while other parameters are kept to our reference values. For very large magnitudes of $b$ the tension actually vanishes and one should obtain acceptable solutions even when reaching the true minimum of the potential. For a large part of the parameter space, the temporary dominance of the effective $\alpha$-potential is a necessary prerequisite for an acceptable solution however.

%
\begin{figure*}[!t] % Quintessence Vs Lambda figures
    \begin{center}
    \begin{psfrags}%
    \psfragscanon%
    %
    % text strings Fig (a):
    \psfrag{f01}[t][t]{\setlength{\tabcolsep}{0pt}
      \begin{tabular}{c}Fig. (a)\end{tabular}}%
    \psfrag{s03}[t][t]{\setlength{\tabcolsep}{0pt}
        \begin{tabular}{c}$z_{\rm J}$\end{tabular}}%
    \psfrag{s04}[b][b]{\setlength{\tabcolsep}{0pt}
        \begin{tabular}{c}$\Delta m_{\rm J}$\end{tabular}}%
    \psfrag{o01}[l][l]{\small{$\Omega_{\rm tot} = 1.00$}}%
    \psfrag{o02}[r][r]{\small{$\Omega_{\Lambda} = 0.8$}}%
    \psfrag{o03}[r][r]{\small{$\Omega_{\Lambda} = 0.70$}}%
    \psfrag{o04}[r][r]{\small{$\Omega_{\Lambda} = 0.6$}}%
    \psfrag{V}[l][l]{\small{$V(\chi)$}}%
    %
    % text strings Fig (b):
    \psfrag{f02}[t][t]{\setlength{\tabcolsep}{0pt}
      \begin{tabular}{c}Fig. (b)\end{tabular}}%
    \psfrag{s05}[t][t]{\setlength{\tabcolsep}{0pt}
        \begin{tabular}{c}$\Omega_{\Lambda}$\end{tabular}}%
    \psfrag{s06}[b][b]{\setlength{\tabcolsep}{0pt}
        \begin{tabular}{c}$\chi^2(\sigma_i)$\end{tabular}}%
    \psfrag{o05}[l][l]{\small{$\Omega_{\Lambda} = 0.73$}}%
    \psfrag{c01}[r][r]{\small{$99\%$ CL}}%
    \psfrag{c02}[r][r]{\small{$95\%$ CL}}%
    \psfrag{c03}[r][r]{\small{$68\%$ CL}}%
    %
    % xticklabels deltaMagn:
	\psfrag{z01}[t][t]{$0$}%
	\psfrag{z02}[t][t]{$0.5$}%
	\psfrag{z03}[t][t]{$1$}%
	\psfrag{z04}[t][t]{$1.5$}%
	\psfrag{z05}[t][t]{$2$}%
    %
    % yticklabels deltaMagn:
	\psfrag{v01}[r][r]{$-0.2$}%
	\psfrag{v02}[r][r]{$-0.1$}%
	\psfrag{v03}[r][r]{$0$}%
	\psfrag{v04}[r][r]{$0.1$}%
	\psfrag{v05}[r][r]{$0.2$}%
    %
    % xticklabels chisquared:
	\psfrag{x01}[t][t]{$0.6$}%
	\psfrag{x02}[t][t]{$0.65$}%
	\psfrag{x03}[t][t]{$0.7$}%
	\psfrag{x04}[t][t]{$0.75$}%
	\psfrag{x05}[t][t]{$0.8$}%
    %
    % yticklabels chisquared:
	\psfrag{y01}[r][r]{$0$}%
	\psfrag{y02}[r][r]{$20$}%
	\psfrag{y03}[r][r]{$40$}%
	\psfrag{y04}[r][r]{$60$}%
	\psfrag{y05}[r][r]{$80$}%
	\psfrag{y06}[r][r]{$100$}%
    %
    % Figure:
    \includegraphics[width=15cm]{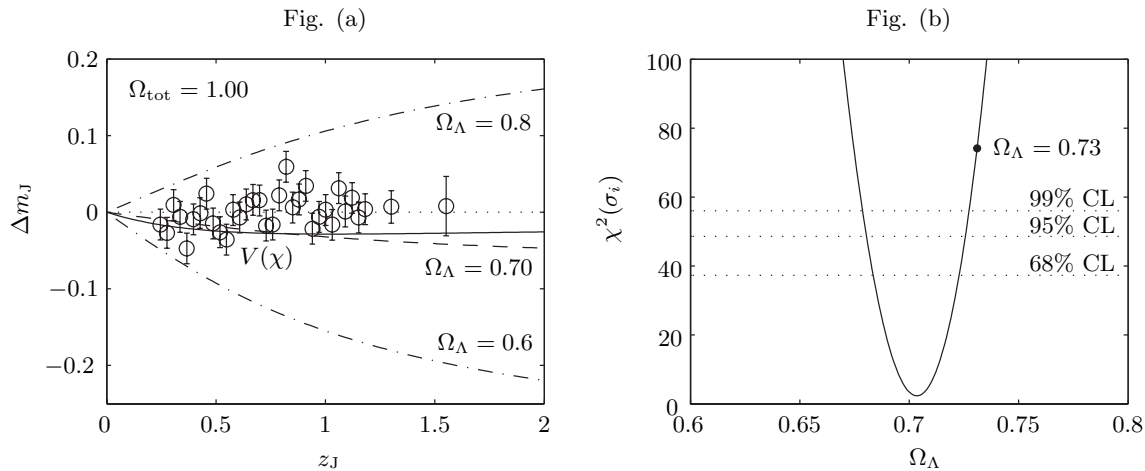}%
    \end{psfrags}%
    \end{center}
    \caption{\small{(a) Observed redshift-magnitude diagram normalized
        to a flat, $\Omega_{\Lambda} = 0.73$ cosmology, with binned
        simulated SNAP data~\cite{SNAPdata}. The solid line representing
        the quintessence-like scenario corresponds to $\Omega_{\chi} = 0.73$
        at present. (b) Statistical $\chi^2(\sigma_i)$ for a cosmological
        constant, using the quintessence-like scenario with
        $\Omega_{\chi} = 0.73$ at present as null hypothesis.}}
    \label{fig:quintLambda}
\end{figure*}
The evolution of both $w_{\chi}$ and $w_{\rm tot}$ are  shown in Fig.~(\ref{fig:w}). The epochs of kination ($w_{\rm tot} = 1$), radiation dominance ($w_{\rm tot} = 1/3$), matter dominance ($w_{\rm tot} = 0$), and the eventual cosmological constant dominance when $w_{\rm tot} \to -1$ are clearly visible.  The evolution of $w_{\chi}$ is much less dramatic. It remains close to unity all the way until $\log_{10}{R} \approx -2$, when it rapidly falls from $w_{\chi} = 1$ to close to $w_{\chi} = -1$, as the kinetic energy of the field becomes neglible in comparision to the potential energy. The only interesting feature is the little bump in $w_{\chi}$ which is caused by the small temporary increase in the kinetic energy of $\rho_{\chi}$, as $\chi$ makes its last roll closer to the potential minimum.

In summary, it is clear from Figs.~(\ref{fig:RtJordan}) and (\ref{fig:rho}) and Table~\ref{table:startPresent} that, except at very early times, the Einstein and Jordan frame are descriptively equivalent and the evolution of all densities have the desired features. In particular, the universe is flat with $\Omega_{\chi} = 0.73$ at present. Also, as shown in Figs.~(\ref{fig:Mbr_alpha}a) and (\ref{fig:Mbr_alpha}b), both $\varphi$ and $\alpha$ fulfill the observational constraints. Overall, the evolution of the energy densities is qualitatively surprisingly similar to the results of Albrecht \emph{et al.}~\cite{albrecht02_2}, apart from the additional evolution at late times connected with the mismatch between the zero of the coupling $\alpha(\chi)$ (\ref{eq:alpha}) and the minimum of the potential $V(\chi)$.

\subsection{Quintessence Vs. $\Lambda$-Cosmology}

Although the scalar field behaves like a cosmological constant at present, it obviously had a varying equation of state in the past. This allows for a possibility to put bounds on the model based on observations that can probe the equation of state at different times. Moreover, the proper way to confront a scalar-tensor theory with supernovae observations is to compute its corresponding magnitude-redshift relation. This focus on the observed expansion rate alone and does not depend on some particular parametrization of the theory. We have therefore computed the magnitude-redshift relation of type Ia supernovae, predicted by our reference case. The result is given in Fig.~(\ref{fig:quintLambda}a), where the magnitude-redshift relation is shown normalized to a corresponding flat, $\Omega_{\Lambda} = 0.73$ cosmology. The model easily fits current data~\cite{SNedata}. However, the magnitude-redshift relation will be  measured to high accuracy by the SNAP-satellite~\cite{SNAP} in the near future. Comparing with the shown binned simulated SNAP data obtained from ref.~\cite{SNAPdata}, suggests that the satellite experiment will be able to distinguish the models from each other. This interpretation is verified by Fig.~(\ref{fig:quintLambda}b), which shows a $\chi^2(\sigma_i)$ function in the space of constant $\Lambda$ models using our quintessence-like scenario with $\Omega_{\chi_0} = 0.73$ as null hypothesis. The $\Omega_{\Lambda}=0.73$ model lies clearly several sigmas away from the present model. However, the minimum of the $\chi^2$-function at $\Omega_{\Lambda} = 0.70$ lies well within one sigma from the null hypothesis. Thus, SNAP data alone will not able to distinguish between our model and the $\Lambda$ models with about 5\% less dark energy.  While the difference is small, it might be possible to break the degeneracy by combining SNAP-data with current data on the cosmic microwave background anisotropies.

\begin{figure*}[!t] % Evolution figures with huge initial field velocity
    \begin{center}
    \begin{psfrags}%
    \psfragscanon%
    %
    % text strings Fig (a):
    \psfrag{f01}[t][t]{\setlength{\tabcolsep}{0pt}
      \begin{tabular}{c}Fig. (a)\end{tabular}}%
    \psfrag{s03}[t][t]{\setlength{\tabcolsep}
        {0pt}\begin{tabular}{c}$\log_{10} R$\end{tabular}}%
    \psfrag{s04}[b][b]{\setlength{\tabcolsep}{0pt}
        \begin{tabular}{c}$\log_{10} \rho$\end{tabular}}%
    %
    % text strings Fig (b):
    \psfrag{f02}[t][t]{\setlength{\tabcolsep}{0pt}
      \begin{tabular}{c}Fig. (b)\end{tabular}}%
    \psfrag{s05}[t][t]{\setlength{\tabcolsep}{0pt}
       \begin{tabular}{c}$\log_{10}{R}$\end{tabular}}%
    \psfrag{s06}[b][b]{\setlength{\tabcolsep}{0pt}
       \begin{tabular}{c}$\varphi/\log{10}$\end{tabular}}%
    \psfrag{s07}[t][b]{\setlength{\tabcolsep}{0pt}
       \begin{tabular}{c}$k^2(\varphi)$\end{tabular}}%
    %
    % xticklabels:
    \psfrag{x01}[t][t]{$-12$}%
    \psfrag{x02}[t][t]{$-7$}%
    \psfrag{x03}[t][t]{$-2$}%
    \psfrag{x04}[t][t]{$3$}%
    \psfrag{x05}[t][t]{$8$}%
    %
    % Densities yticklabels:
    \psfrag{v01}[r][r]{$-160$}%
    \psfrag{v02}[r][r]{$-140$}%
    \psfrag{v03}[r][r]{$-120$}%
    \psfrag{v04}[r][r]{$-100$}%
    \psfrag{v05}[r][r]{$-80$}%
    \psfrag{v06}[r][r]{$-60$}%
    %
    % Mbr yticklabels:
    \psfrag{r01}[r][r]{$14$}%
    \psfrag{r02}[r][r]{$14.5$}%
    \psfrag{r03}[r][r]{$15$}%
    \psfrag{r04}[r][r]{$15.5$}%
    \psfrag{r05}[r][r]{$16$}%
    %
    % k^2 yticklabels:
	\psfrag{y01}[l][l]{$0$}%
	\psfrag{y02}[l][l]{$2$}%
	\psfrag{y03}[l][l]{$4$}%
	\psfrag{y04}[l][l]{$6$}%
	\psfrag{y05}[l][l]{$8$}%
	\psfrag{y06}[l][l]{$10$}%
	\psfrag{y07}[l][l]{$12$}%
    \psfrag{ypower1}[Bl][Bl]{$10^{19} \times$}%
    %
    % Figure:
    \includegraphics[width=15cm]{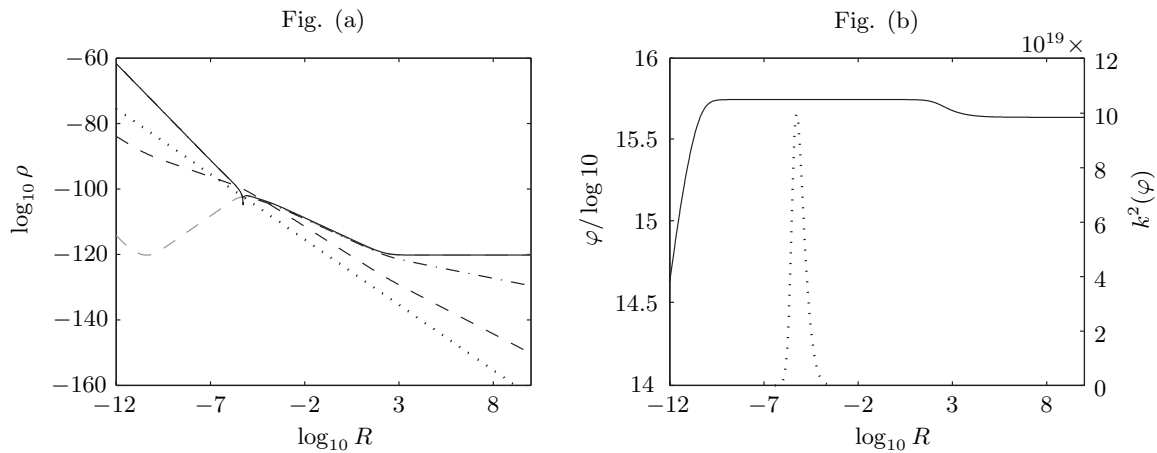}%
    \end{psfrags}%
    \end{center}
    \caption{\small{(a) Shown are energy densities (in Planck units) as a
        function of the Einstein frame scale factor $R$ for the same model
        parameters and initial conditions as in Fig.~(\ref{fig:rho}), except
        for the initial field velocity which now has been set to
        $\partial \chi / \partial t = 2.0 \times 10^{-31}$ $M_{\rm b}^2$.
        Different curves represent matter (dashed black), radiation (dotted),
        total scalar energy density (solid), scalar potential energy
        (grey dashed) and scalar kinetic energy (dot-dashed). (b) Shown is
        $\varphi/\log{10}$ ($= \log_{10} M_{\rm b}r$) (solid line) and the
        $k^2(\varphi)$-factor defined in Eqn.~(\ref{eq:ksquared})
        (dotted line) as a function of the Einstein frame scale factor $R$.}}
    \label{fig:hugeKin}
\end{figure*}
Let us finally remember that the late time evolution, due to the tension between the zero of the coupling $\alpha$ and the minimum of the potential $V$, is a rather generic feature of the present model, and therefore so is the observed difference in the magnitude-redshift relation between the present model and a corresponding constant $\Lambda$ model.

\subsection{Robustness of the New Stabilization Mechanism}

Finally, let us consider the robustness of the new stabilization mechanism based on the $A = 0$ singularity.  On one hand, the barrier height at the singularity is tiny compared to the value of the potential when $\varphi \approx 0$, but on the other, it is about 20 orders of magnitude higher than the value of the potential at the minimum. Could it nevertheless be possible to go past the barrier and make the world enter the region where gravitons would carry negative energy in the Jordan frame by giving the field a much larger initial velocity? 
The answer turns out to be negative, for all initial velocity we gave the field, the solution always either got stuck close to the singularity, or got its kinetic energy diluted by the Hubble friction.

Figs.~(\ref{fig:hugeKin}a--\ref{fig:hugeKin}b) shows an example of an evolution with huge initial field velocity, where we have used the same reference values of the model parameters and initial conditions as shown in Table~\ref{table:startPresent}, apart from giving the field 12 orders of magnitude larger kinetic energy.  Now the potential reaches its minimum already at $\log_{10}{R}Ê\approx -11$ and starts increasing again as the field continues climbing the potential barrier. The growth of $\chi$ and therefore also that of $\varphi$ stops at $\log_{10}{R}Ê\approx -6$, where the potential energy reaches its maximum and the kinetic energy goes through zero (this shows up as a small dip in Fig.~(\ref{fig:hugeKin}a)).  The huge spike in the $k^2(\varphi)$-function shows that the field came close to the singularity before turning back. As the field rolls back, $\varphi$ is changing slowly initially, because of the near singular relation between $\chi$ and $\varphi$, but eventually  at about $\log_{10}{R}Ê\approx 3$ (far in the future) $\varphi$ makes  a small adjustment before freezing out completely. The field then rolls back towards the minimum as the Hubble friction absorbs the remaining kinetic energy. Observe that most of the time during the period of backwards rolling the universe is neither matter, radiation, or cosmological constant dominated, nor does it behave like during $\dot\chi^2$-dominated kination. Instead, it is a peculiar mix of kinetic ($k$-essence) and potential (quintessence) energy. Needless to say, this example is entirely incompatible with our universe; we only gave it to show the robustness of the radius stabilization against initial conditions with very large field velocities.

% New Section
%%%%%%%%%%%%%%%%%%%%%%%%%%%%%%%%%%%%%%%%%%%%%%%%%%%%%%%%%%%%%%%%%%%%%%%%%%
%
%  --------------------------- Section 8 ------------------------------
%
%%%%%%%%%%%%%%%%%%%%%%%%%%%%%%%%%%%%%%%%%%%%%%%%%%%%%%%%%%%%%%%%%%%%%%%%%%

\section{Summary and Discussion}
\label{sect:summary}

We have considered in detail the cosmologies of a class of dilatonic scalar-tensor theories of gravity inspired by large extra dimensions and based on ideas put forward by Albrecht, Burgess, Ravndall and Skordis~\cite{albrecht02_1,albrecht02_2}. In the LED context, the effective quintessence field $\chi$ corresponds to a radion mode originating from dimensional reduction of a full $(4+d)$-dimensional action to an effective 4D brane action. We have shown that observational gravity constraints (the coupling $\alpha$) require that radiative corrections from additional bulk fields $\phi_i$ nearly cancel the tree level contribution to the effective gravitational constant $G_{\rm eff}$ at present. The corresponding zero in the scaling function $A(\varphi)$ of the Ricci scalar in the Jordan frame, shows up as a singularity in the effective potential in the Einstein frame, yielding a new type of stabilization for the field $\varphi$ (i.e. the radius in the LED context). In this stabilization scheme the model is thus (almost) automatically consistent with the gravity constraints. If this mechanism is implemented in a perturbatively converging scheme, the lowest order corrections to the Casimir potential are truly small, and can not give rise to a stabilization, as was assumed in the original work~\cite{albrecht02_2}.

Depending on the parameters, the model can give rise to features very different from a generic quintessence model. To a degree this was already recognized by the earlier work, and indeed the model has been dubbed ``walking quintessence"~\cite{burgess02}.  However, we have shown that the model can behave at times more like $k$-essence, where the dynamics is mainly controlled by the kinetic term, or even as a phantom cosmology, where the scalar field has negative kinetic energy. Using perfectly natural values of the model parameters, it is even possible to arrange a phantom phase that has finite range in $\varphi$, which might allow for a finite duration phantom phase, possibly side-stepping the associated problems with vacuum instability in such models~\cite{phantom}. However, whether such a region can be reached dynamically in a model compatible with the observational constraints is yet unknown to us; although it seems unlikely in the subset of models studied in this paper, the possibility at large remains an open question.

Imposing various constraints very nearly determine all the fundamental parameters in the effective 4D brane action. Moreover, some of the parameters or combinations of parameters needs to be fine tuned for the model to work. (That is, to simultaneously give rise to the observed strength of gravity and obey the $\alpha$-bound.) Nevertheless, having fixed the parameters, we show that the model can give rise to an acceptable cosmological evolution. Moreover, the model predicts a characteristic signature in the magnitude-redshift relation of type Ia supernovae, that may make it possible for a combination of supernova and CMBR data to distinguish the model from a pure $\Lambda$-cosmology. When applying the observational constraints, we made the assumption that the behaviour of our model should be very close to General Relativity. Although beyond the scope of this paper, a complete observational analysis requires a computation of the corresponding CMBR spectrum and the cosmological perturbations constrained by large scale structures.

A particularily attractive feature of the present model is that it makes a clear identification for the origin of the effective quintessence field in a well defined context that has other motivations beyond the cosmological application. Indeed, models with large extra dimensions were originally proposed as a novel solution to the heirarchy problem~\cite{Hamedetal,Antoniadetal}. This blessing comes with a burden however. We have seen that cosmology puts several stringent constraints on the model parameters and leaves little freedom in choosing the initial conditions of the cosmological evolution. A particularily unappealing feature in this context is that one is forced to assume a very large value for the radius $r$ at the time when the universe enters the radiation dominance. Indeed, the thermal history in LED models is heavily constrained by the need to avoid overproducing Kaluza-Klein modes, and not much can be said with a certainty about the conditions prevailing at or before the reheating phase. We did not discuss these issues in any detail in this work, and refer the reader to the original papers~\cite{albrecht02_1,albrecht02_2} for elaboration in this matter.

While it is certainly rewarding to be able to put the model into a larger theoretical framework, the scalar-tensor theory of Eqn.~(\ref{eq:S}) can indeed be viewed as the starting point. In this context the parametrization (\ref{eq:SJordan}--\ref{eq:C}) emerges as a simple correction to the usual dilatonic scalar-tensor gravity. Adopting this point of view the model becomes much less severely constrained, and one could try also more complicated functional forms for the scaling functions $A$, $B$ and $C$, for example to see if one could more easily construct phantom-like cosmologies consistent with all astrophysical observations.

%%%%%%%%%%%%%%%%%%%%%%%%%%%%%%%%%%%%%%%%%%%%%%%%%%%%%%%%%%%%%%%%%%%%%%%%%%
%
%  ------------------------- Acknowledgements ----------------------------
%
%%%%%%%%%%%%%%%%%%%%%%%%%%%%%%%%%%%%%%%%%%%%%%%%%%%%%%%%%%%%%%%%%%%%%%%%%%

\begin{acknowledgments}
We would like to thank Cliff Burgess for very helpful correspondence on 
aspects of the LED induced effective action and Edward M\"ortsell for useful discussions on supernovae constraints.

This work was partially supported by the CIMO-grant TM-04-2500.
\end{acknowledgments}

%%%%%%%%%%%%%%%%%%%%%%%%%%%%%%%%%%%%%%%%%%%%%%%%%%%%%%%%%%%%%%%%%%%%%%%%%%
%
%  --------------------------- Appendix A --------------------------------
%
%%%%%%%%%%%%%%%%%%%%%%%%%%%%%%%%%%%%%%%%%%%%%%%%%%%%%%%%%%%%%%%%%%%%%%%%%%

\appendix

\section{Explicit Relation\protect\\Between $\chi$ and $\varphi$}
\label{appendixA}

\noindent The defining relation for the field $\chi$ is:
\begin{equation}
  (\partial \chi)^{2} \equiv
    M_{\rm b}^{2}
       \left[
         \frac{3}{2} \left(2 + \frac{a}{A}\right)^2 - \frac{2B}{A}
       \right]
    (\partial \varphi)^{2}
\label{eq:Xdefvarphi}
\end{equation}
where $A = 1 + a\varphi$, $B= 1 + b\varphi$ and $C= 1 + c\varphi$. The function appearing in the square brackets can be written as a polynominal in $\widetilde{A} \equiv A/a$:
\begin{eqnarray}
  \mathbb{R} & = & \frac{3}{2}(2A + a)^2 - 2AB \nonumber \\
  & \equiv & \alpha + \beta \widetilde{A} + \gamma \widetilde{A}^2 \: ,
\end{eqnarray}
where
\begin{equation}
  \alpha = \frac{3 a^2}{2},  \quad
  \beta =  2(3a^2 - a + b),  \quad
  \gamma = 2a (3a - b) \: ,
\end{equation}
and we define
\begin{equation}
  \Delta \equiv 4 \alpha \gamma - \beta^2 \: .
\end{equation}
Rewriting Eqn.~(\ref{eq:Xdefvarphi}) solely in terms of $\tilde A$:
\begin{equation}
   (\partial \chi)^2 \equiv M_{\rm b}^2
      \frac{\mathbb{R}}{a^2\widetilde{A}^2}
      (\partial \widetilde{A})^2 \: .
\label{XdefAtilde}
\end{equation}
The solution to Eqn.~(\ref{XdefAtilde}) can be found by direct integration. For $\beta$, $\gamma$, $\Delta \neq 0$ one finds
\begin{eqnarray}
   \frac{\chi}{M_{\rm b}}
   &=& \frac{1}{|a|}
      \Bigg( \sqrt{\mathbb{R}} + \frac{\beta}{2}\mathbb{X} \\Ê\nonumber
   & & \phantom{\frac{1}{|a|} \Bigg(}
      {}- \sqrt{\alpha}
         \log{ \left| \frac{2 \sqrt{\alpha \mathbb{R}}
                      + \beta \widetilde{A}
                      + 2 \alpha }{\widetilde{A}}
              \right| } \Bigg)
         \times \textrm{sgn} \, \widetilde{A} \: ,
\end{eqnarray}
where
\begin{equation}
  \mathbb{X} =
  \left\{
    \begin{array}{ll}
        \frac{1}{\sqrt{\gamma}}
           \log{\left| 2 \sqrt{\gamma \mathbb{R}}
                      + 2 \gamma \widetilde{A} + \beta
               \right|}
                & \textrm{for $\gamma > 0$} \\
                & \\
        \frac{-1}{\sqrt{-\gamma}}
          \arcsin{  \frac{2 \gamma \widetilde{A} +
                    \beta }{\sqrt{-\Delta}} }
               & \textrm{for $\gamma < 0$}
        \end{array} \: .
  \right.
\end{equation}
If $\beta = 0$ the solution is instead
\begin{equation}
  \frac{\chi}{M_{\rm b}} = \frac{1}{|a|} \left(
  \sqrt{\mathbb{R}} - \sqrt{\alpha} \log{\left| \frac{
                \sqrt{\mathbb{R}} + \sqrt{\alpha} }{
                \widetilde{A}} \right| }
        \right) \times \textrm{sgn} \, \widetilde{A} \: ,
\end{equation}
for $\gamma = 0$ one finds
\begin{equation}
  \frac{\chi}{M_{\rm b}}
       = \frac{1}{|a|}
        \left( 2 \sqrt{\mathbb{R}} + \sqrt{\alpha}
                \log{\left| \frac{
                  \sqrt{\mathbb{R}} - \sqrt{\alpha} }{
                  \sqrt{\mathbb{R}} + \sqrt{\alpha}} \right| }
        \right) \times \textrm{sgn} \, \widetilde{A} \: ,
\end{equation}
and finally for $\Delta = 0$,
\begin{eqnarray}
   \frac{\chi}{M_{\rm b}} &=&
      \frac{1}{|a|} \bigg(\sqrt{\gamma} \widetilde{A}
         + \sqrt{\alpha} \log{| \widetilde{A} |} \bigg) \\ \nonumber
       & & \times \, \textrm{sgn} \left( \sqrt{\gamma} +
             \sqrt{\alpha} / \widetilde{A} \right) \: .
\end{eqnarray}
All of the above solutions are valid in the range $-0.1 < a < 0.1$ and $-0.1 < b < 0.1$, given that they $a$Êand $b$Êare arranged so that $k^2(\varphi) > 0$.

%%%%%%%%%%%%%%%%%%%%%%%%%%%%%%%%%%%%%%%%%%%%%%%%%%%%%%%%%%%%%%%%%%%%%%%%%%
%
%  --------------------------- Bibliography ------------------------------
%
%%%%%%%%%%%%%%%%%%%%%%%%%%%%%%%%%%%%%%%%%%%%%%%%%%%%%%%%%%%%%%%%%%%%%%%%%%

\end{document}